\documentclass[useAMS, usenatbib, usegraphicx]{mn2e}

\usepackage{hyperref}
\usepackage[T1]{fontenc}
\usepackage{aecompl}

\newcommand{\despotic}{\texttt{DESPOTIC}}
\newcommand{\muH}{\mu_{\rm H}}
\newcommand{\mH}{m_{\rm H}}

\begin{document}

\title[DESPOTIC]{DESPOTIC -- A New Software Library to Derive the Energetics and SPectra of Optically Thick Interstellar Clouds}

\author[Krumholz]{
Mark R. Krumholz$^1$\thanks{mkrumhol@ucsc.edu}
\\ \\
$^1$Department of Astronomy \& Astrophysics, University of California, Santa 
Cruz, CA 95064 USA}

\maketitle

\begin{abstract}
I describe \despotic, a code to Derive the Energetics and SPectra of Optically Thick Interstellar Clouds. \despotic\ represents such clouds using a one-zone model, and can calculate line luminosities, line cooling rates, and in restricted cases line profiles using an escape probability formalism. It also includes approximate treatments of the dominant heating, cooling, and chemical processes for the cold interstellar medium, including cosmic ray and X-ray heating, grain photoelectric heating, heating of the dust by infrared and ultraviolet radiation, thermal cooling of the dust, collisional energy exchange between dust and gas, and a simple network for carbon chemistry. Based on these heating, cooling, and chemical rates, \despotic\ can calculate clouds' equilibrium gas and dust temperatures, equilibrium carbon chemical state, and time-dependent thermal and chemical evolution. The software is intended to allow rapid and interactive calculation of clouds' characteristic temperatures, identification of their dominant heating and cooling mechanisms, and prediction of their observable spectra across a wide range of interstellar environments. \despotic\ is implemented as a Python package, and is released under the GNU General Public License.
\end{abstract}


\section{Introduction}
\label{sec:intro}

With the advent of powerful radio telescopes such as the Atacama Large Millimeter Array (ALMA), it has become possible to study the cold interstellar medium (ISM) in unprecedented detail and at greater distances than ever before. Observations from these facilities have stimulated a great deal of theoretical interest in the properties of the cold ISM, both nearby and in environments far-removed from those found near the Sun. One of the goals of these theoretical investigations has been to study how the thermodynamics of gas, and thus the nature of the star formation process within it, varies as a function of environment. A second goal has been to predict the observable emission of gas in a variety of environments.

Theoretical investigations of this sort often benefit from approximate calculations using idealized geometries that can produce relatively fast results, while also including a wide range of microphysical processes in order to determine which ones are important. However, there are few publicly-available tools capable of performing these functions for the dense, optically thick phase of the interstellar medium. Traditional photodissociation region (PDR) codes \citep[e.g.][]{meijerink05a, le-petit06a}, or codes that can handle a variety of ISM phases such as \texttt{cloudy} \citep{ferland98a}, perform calculations in this regime, but the complexity of the problem means that these codes are too computationally-costly for either broad surveys or quick, interactive scans of parameter space. A number of authors have released codes capable of performing fast calculations of molecular emission line spectra using the large velocity gradient or various other forms of the escape probability approximation (e.g.~\texttt{CASSIS}\footnote{\url{http://cassis.irap.omp.eu/}} and \texttt{RADEX}, \citealt{van-der-tak07a}). While these are useful tools for the analysis of observations, they are only capable of predicting line emission given fixed physical conditions, and they do not calculate many quantities of interest for theoretical modeling, such as rates of heating and cooling, thermal equilibria, or time-dependent thermal behavior.

The need for codes that are capable of performing calculations of this sort is apparent from the wide variety of applications they have found in the recent literature. For example, \citet{goldsmith01a} and \citet{lesaffre05a} investigate the temperature structure within protostellar cores. \citet{krumholz07g} use an escape probability model to study the relationship between star formation rates and emission is a variety of molecular lines. \citet{krumholz11b} use thermal equilibrium models of the ISM to explore the relationship between star formation and the chemical state of the gas. \citet{narayanan11a, narayanan12a}, \citet{shetty11a, shetty11b}, and \citet{feldmann12a, feldmann12b} all investigate the conversion between observed CO luminosity and molecular mass using simulations of galaxies, coupled to post-processing to predict the observable line emission. \citet{narayanan12b, narayanan12c} perform calculations of interstellar medium temperatures as a way of estimating the Jeans mass in molecular clouds, and its possible implications for changes in the stellar IMF over cosmological times. \citet{papadopoulos10a} and \citet{meijerink11a} consider star formation in extreme environments with X-ray and cosmic-ray fluxes far higher than are found in the Solar neighborhood, and in the process rely on calculations of the thermal behavior of gas under these conditions. Similarly, \citet{munoz13a} study the ISM in high-redshift galaxies where the metallicity is much lower and the cosmic microwave background is much hotter than in the present-day universe. With a few exceptions, all of these authors developed their own custom codes to model the thermodynamics and line emission of the cold ISM. However, this effort is largely duplicative, since these calculations all involve the same related set of problems. Moreover, the results of the calculations can be difficult to compare due to the differing assumptions and approximations made by the various authors in their modeling, not all of which are well-documented in the literature.

In order to support theoretical investigations facing problems of this sort, reduce duplication of effort, and encourage calculations with documented, open-source tools to allow easy comparison between authors, I have developed a software library to Derive the Energetics and SPectra of Optically Thick Interstellar Clouds (\despotic). \despotic\ uses an escape probability formalism to calculate line emission, and couples this to a calculation of either equilibrium or time-dependent gas and dust temperatures, including the dominant processes in a wide variety of environments: cosmic-ray and X-ray ionization heating, photoelectric heating, grain-gas energy exchange, and radiative heating and cooling of dust grains. The software is implemented as Python package, enabling easy, interactive calculation, and also easy integration with other software. It also provides an automated interface with the Leiden Atomic and Molecular Database (LAMDA; \citealt{schoier05a}). \despotic\ is publicly available from a \href{https://sites.google.com/a/ucsc.edu/krumholz/codes/despotic}{\texttt{dedicated web page}}, from \href{https://code.google.com/p/despotic/source/checkout}{\texttt{google code}}, and from the \href{https://pypi.python.org/pypi/DESPOTIC/1.1.0}{\texttt{Python Package Index}}, and is released under the GNU General Public License. It comes with extensive documentation, including a User's Guide with a full listing of all routines and options.

In the remainder of this paper, I describe the model system that \despotic\ uses and the equations it can solve (\S~\ref{sec:model}) and the numerical methods by which it solves those equations (\S~\ref{sec:methods}). I then provide some example applications (\S~\ref{sec:examples}), provide some warnings about the limitations of the code (\S~\ref{sec:limitations}), and summarize (\S~\ref{sec:summary}).

\section{Model System and Equations Solved}
\label{sec:model}

\subsection{Physical Model}

The basic physical system treated by \despotic\ is a uniform spherical cloud (though other simple geometries are provided as options, as described below). Such a cloud is characterized by several physical and chemical properties, which are taken to be uniform unless stated otherwise. The physical properties are a volume density of hydrogen nuclei $n_{\rm H}$ and a mean column density of hydrogen nuclei $N_{\rm H}$, the gas temperature $T_g$, the dust temperature $T_d$, the non-thermal velocity dispersion $\sigma_{\rm NT}$, and (optionally) a bulk radial velocity gradient $dv_r/dr$. Note that \despotic\ defines the column density as an average over the cloud, i.e.~it is the total number of hydrogen atoms in the cloud divided by the cloud's cross-sectional area.

The dust within a cloud is characterized by six quantities. Three of these describe the dust cross-section per H nucleus to thermal radiation at temperature T = 10 K, $\sigma_{d,10}$, to radiation in the range of $8 - 13.6$ eV that dominates photoelectron production, $\sigma_{d,\rm PE}$, and averaged over the diffuse interstellar radiation field (ISRF) $\sigma_{d,\rm ISRF}$. The fourth quantity is the total dust abundance normalized to the Milky Way value, $Z_d'$. The remaining two quantities are the dust spectral index $\beta$ for thermal radiation, and the gas-grain collisional coupling coefficient $\alpha_{\rm GD}$. I define all of these terms in detail below.

\despotic\ parameterizes the radiation field (including cosmic rays) around the cloud by the following quantities: $\zeta$ gives the primary ionization rate per H nucleus due to hard x-ray photons and cosmic rays, $\chi$ describes the energy density, normalized to the Solar neighborhood value, of the non-thermal interstellar radiation field produced primarily by stars, $T_{\rm rad,dust}$ gives the infrared radiation field seen by the dust, and $T_{\rm CMB}$ is the cosmic microwave background temperature.

Finally, the chemical composition of the cloud is described by the abundances of bulk constituents and trace emitting species. The abundances of the bulk constituents in the \despotic\ model are given by $x_{\rm HI}$, $x_{\rm pH_2}$, $x_{\rm oH_2}$, $x_{\rm He}$, $x_{\rm e}$, and $x_{\rm H^+}$, which describe atomic hydrogen, para-H$_2$, ortho-H$_2$, helium, free electrons, and free protons, respectively.\footnote{Although \despotic\ includes free protons and electrons, it is only intended for use in regions where the gas is predominantly neutral, i.e.~ $x_{\rm H^+} \ll x_{\rm HI} + 2 (x_{\rm pH_2} + x_{\rm oH_2})$, and similarly for $x_e$. It does not include many heating and cooling processes that are important in highly ionized regions.} The abundances of emitting species (e.g.~CO, HCN, H$_2$O, etc.) are characterized in the same way, with $x_i$ representing the abundance of the $i$th emitting species.

Given the bulk composition, one can also compute a number of additional quantities, of which we will make use below. Three of these are the mean mass per H nucleus $\muH$, the mean mass per free particle $\mu$, and the isothermal sound speed $c_s$, given by
\begin{eqnarray}
\muH & = & x_{\rm HI} + x_{\rm H^+} + 2(x_{\rm pH_2}+x_{\rm oH_2}) + 4 x_{\rm He} \\
\mu & = & \frac{\muH}{x_{\rm HI} + x_{\rm H^+} + x_{\rm pH_2}+x_{\rm oH_2} + x_{\rm He} + x_{\rm e}} \\
c_s & = & \sqrt{k_B T_g/\mu \mH},
\end{eqnarray}
where $\muH$ and $\mu$ are measured in units of the hydrogen mass $\mH$. Note this this expression neglects the mass of electrons, and assumes that emitting species contribute negligibly to the mass. Two additional quantities are the gas specific heat at constant volume $c_{v, \rm H}$ and at constant pressure $c_{p,\rm H}$, which for convenience we express per H nucleus rather than per unit mass or per unit volume. Thus $c_{v,\rm H}$ and $c_{p,\rm H}$ have units of energy over temperature, and can be converted to the usual values per unit mass simply by multiplying by a $\mu_{\rm H} m_{\rm H}$. Calculation of the specific heats requires some care when the chemical composition includes molecular hydrogen. I discuss this topic in detail, and derive \despotic's expressions for $c_{v,\rm H}$ and $c_{p, \rm H}$, in Appendix \ref{sec:cv}.

The final quantity one can compute is the clumping factor $f_{\rm cl}$ for the cloud, which represents an enhancement in the rates of all collisional processes due to non-uniformity of the gas. The quantity $n_{\rm H}$ is the volume-averaged density over the cloud, but in a non-uniform cloud the density $n_{\rm H}(\bmath{x})$ at any position $\bmath{x}$ may be higher or lower than this. Since the rate of collisions per unit volume at a given position varies as $n_{\rm H}(\bmath{x})^2$, the rate of collisions per H atom in a non-uniform cloud exceeds that in a uniform cloud by a factor
\begin{equation}
f_{\rm cl} = \frac{\langle n_{\rm H}(\bmath{x})^2 \rangle}{n_{\rm H}},
\end{equation}
where the angle brackets indicate an average over the cloud volume; thus $f_{\rm cl}$ is simply the factor by which the mass-weighted mean density exceeds the volume-weighed mean density. For a supersonically turbulent medium, this factor is approximately (\citealt{ostriker01a, padoan02a}; also see \citealt{lemaster08a}, \citet{federrath08a}, and \citealt{price11a})
\begin{equation}
f_{\rm cl} \approx \sqrt{1 + 0.75 \sigma_{\rm NT}^2/c_s^2}.
\end{equation}

\subsection{Heating and Cooling Processes}

The gas heating and cooling processes included in \despotic\ are ionization heating, heating by the grain photoelectric effect, gravitational compression heating, line cooling, and either heating or cooling by collisional energy exchange between dust and gas. The grain heating and cooling processes included in \despotic\ are cooling by thermal radiation, heating by the interstellar radiation field, heating by an infrared radiation field, heating by the cosmic microwave background radiation, heating by absorption of line radiation, and collisional coupling to the gas. 

Given this list of processes, the time rate of change of the gas energy per H nucleus $e_{g,\rm sp}$ as
\begin{equation}
\label{eq:dedtgas}
\frac{de_{g, \rm sp}}{dt} = \Gamma_{\rm ion} + \Gamma_{\rm PE} + \Gamma_{\rm grav} - \Lambda_{\rm line} + \Psi_{\rm gd},
\end{equation}
where $\Gamma_{\rm ion}$, $\Gamma_{\rm PE}$, and $\Gamma_{\rm grav}$ are the rates of ionization, photoelectric, and gravitational heating per H nucleus, $\Lambda_{\rm line}$ is the rate of line cooling per H nucleus, and $\Psi_{\rm gd}$ is the rate of dust-gas energy exchange per H nucleus. I give explicit formulae for all these terms in Appendix \ref{sec:heatcool}. The corresponding time rate of change of the temperature is
\begin{equation}
\label{eq:tempevol}
\frac{dT_g}{dt} = \frac{1}{(c_{v,\rm H},c_{p,\rm H})}\left(\Gamma_{\rm ion} + \Gamma_{\rm PE} + \Gamma_{\rm grav} - \Lambda_{\rm line} + \Psi_{\rm gd}\right),
\end{equation}
where $c_{v,\rm H}$ is the gas specific heat per H nucleus at constant volume and $c_{p,\rm H}$ is the specific heat per H nucleus at constant pressure (which are calculated in Appendix \ref{sec:cv}). The parentheses indicate that one can use either $c_{v,\rm H}$ or $c_{p,\rm H}$ in the above equation, depending on whether one wishes to consider gas cooling isochorically or isobarically.

Similarly, for the dust grains the total rate of change of specific energy per H nucleus is
\begin{equation}
\label{eq:dedtdust}
\frac{de_{d,\rm sp}}{dt} = \Gamma_{\rm ISRF} + \Gamma_{d, \rm line} + \Gamma_{d, \rm CMB} + \Gamma_{d,\rm IR} - \Lambda_{d} - \Psi_{\rm gd},
\end{equation}
where $\Gamma_{\rm ISRF}$, $\Gamma_{d,\rm line}$, $\Gamma_{d,\rm CMB}$, and $\Gamma_{d,\rm IR}$ are the rates of heating due to the interstellar radiation field, line radiation, the cosmic microwave background, and infrared radiation, and $\Lambda_d$ is the rate of dust cooling by thermal radiation. As with the gas heating and cooling processes, I give explicit formulae for all these terms in Appendix \ref{sec:heatcool}. In principle one could consider time-dependent temperature evolution of the dust as well as of the gas, but since the specific heat of the dust is far less than that of the gas, and is a complex function of the properties of the grains, \despotic\ does not treat this case. Instead, it assumes that the grain population is always in thermal equilibrium.

\subsection{Chemical Processes}

In addition to thermal processes, \despotic\ can also calculate chemical processes that cause the abundances $x_i$ of various species to change with time. \despotic\ allows users to define arbitrary chemical networks by specifying a set of species and a set of chemical reaction rate equations of the form
\begin{equation}
\label{eq:chemeq}
\frac{d\mathbf{x}}{dt} = f(\mathbf{x}, n_{\rm H}, N_{\rm H}, T_g, \zeta, \ldots),
\end{equation}
where $\mathbf{x}$ is the vector of fractional abundances for the various species in the network, and the reaction rates on the right-hand side can be a function of these abundances, of the overall volume density, column density, gas temperature, ionization rate, radiation field, or any of the other quantities that \despotic\ uses to describe a cloud. Once specified, the equations can be integrated over a specified time or until the chemical state reaches equilibrium. The repository version of \despotic\ implements the reduced carbon-oxygen chemistry network of \citet{nelson99a}, which models the processes leading to the transition from C$^+$- to CO-dominated composition in molecular clouds.

\subsection{Line Shapes}

\begin{figure}
\includegraphics[width=84mm]{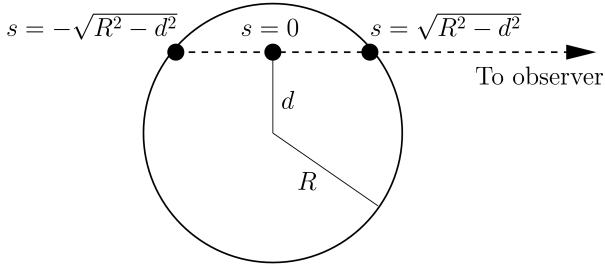}
\caption{
\label{fig:cloudfig}
Diagram of the geometry used by \despotic\ when calculating line shapes. The circle shows the cloud, with radius $R = 3N_{\rm H}/4n_{\rm H}$, and the dashed line is the observer's line of sight through it.
}
\end{figure}

\despotic's final major capability is calculating the profiles of spectral lines. In general this is not a useful calculation in a one-zone escape probability model; since the level populations in such a model are assumed to be uniform, the result is necessarily rather uninteresting, and is simply given by the usual solution to the transfer equation for media with emission and absorption coefficients that are independent of position. However, one can relax the assumption of uniform level populations by making another one: that the species is in LTE, and that the temperature $T$ is a known function of position.\footnote{In principle one in fact needs to know only the excitation temperature $T_{\rm ex}$ for the two levels that produce the line, together with the number density the atoms or molecules that are in the lower state $n_{\ell}$. However, in practice it is unlikely that one will simultaneously know $T_{\rm ex}$ and $n_\ell$ in any situation other than when the levels are in LTE, and thus I limit the discussion to this case. If one does in fact know $T_{\rm ex}$ and $n_\ell$, it is trivial to perform a calculation for that case simply by setting the level populations to their LTE values at $T_{\rm ex}$, and adjusting the overall density of the species so that $n_\ell$ has the desired value.} Solving for the shapes of lines in this limit allows the code to compute pCygni and inverse pCygni profiles, among other applications. This computation is performed for a spherical cloud following \despotic's general model, and consider a line of sight passing through it at an offset distance $d$ from the cloud center (see Figure \ref{fig:cloudfig}). Details of how this calculation is performed are given in Appendix \ref{app:lineshape}.

\section{Code Architecture and Algorithms}
\label{sec:methods}

In this section I describe the architecture of the \despotic\ code and the algorithms it uses to solve the equations introduced in the previous section.

\subsection{Overall Architecture}

\despotic\ is a library intended not only to be used for stand-alone calculations, but also to allow easy extensibility, easy integration with other codes, and to allow users to conduct interactive, exploratory calculations. To this end, \despotic\ is implemented as a Python package, which allows a very high level of abstraction such that many useful computations can be performed with no more than a single line of code on the part of the user. To achieve high performance, \despotic\ makes extensive use of the ability of the \texttt{numPy} and \texttt{sciPy} libraries to interface with the fast, optimized numerical libraries \texttt{LAPACK}\footnote{\url{http://www.netlib.org/lapack/}} \citep{anderson99a}, \texttt{MINPACK}\footnote{\url{http://www.netlib.org/minpack/}} \citep{more80a}, and \texttt{ODEPACK}\footnote{\url{https://computation.llnl.gov/casc/odepack/odepack_home.html}} \citep{hindmarsh83a}. It is hard to provide a quantitative estimate of code execution times for \despotic\ routines, since as I discuss below the most computationally-intensive ones require iterative methods, and the time required for such a solution is a strong function of the quality of the starting guess. Nonetheless, I give a general idea of code execution times, as tested on a single processor of a modern workstation, for some example applications in \S~\ref{sec:examples}. Individual instances of \despotic\ classes use internal private storage, and thus are thread-safe should a user desire to use threading to accelerate the calculation of large grids of models via the standard Python threading interface. Threading of internal \despotic\ calculations for single clouds will be added in a future release.

\subsection{Capabilities and Algorithms}

\subsubsection{Level Populations and Line Luminosities}

The most basic capability of \despotic\ is to compute level populations and line luminosities for an emitting species embedded in a cloud of specified physical properties ($n_{\rm H}$, $T_g$, $\sigma_{\rm NT}$, abundances, etc.). The emitted intensity for any line is given by equation (\ref{eq:intintensity}), and the numerical algorithm for calculating level populations and line luminosities is given in Appendix \ref{sec:lines}. The computation can be performed either assuming the cloud is optically thin, or using the escape probability approximation for an optically thick cloud. Note that this is the same computation performed by codes like \texttt{RADEX} \citep{van-der-tak07a} and \texttt{lineLum} \citep{krumholz07g}, and the latter is the direct ancestor of the corresponding portion of \despotic. Appendix \ref{sec:radex} provides a direct comparison between \despotic\ and \texttt{RADEX}. 

\subsubsection{Cooling Rates, Thermal Equilibria, and Time-Dependent Temperature Evolution}
\label{sec:therm}

In addition to computing line luminosities and level populations, \despotic\ can also compute the heating and cooling rates of gas and dust. It does so by evaluating all the terms in equations (\ref{eq:dedtgas}) and (\ref{eq:dedtdust}); since one of these terms is $\Lambda_{\rm line}$, this procedure entails solving for the level populations and escape probabilities.

\despotic\ can also solve for equilibrium dust and gas temperatures. \despotic\ obtains these values by setting $de_{g,\rm sp}/dt = 0$ in equation (\ref{eq:dedtgas}) and $de_{d,\rm sp}/dt = 0$ in equation (\ref{eq:dedtdust}). The user can also add arbitrary additional heating and cooling terms to either equation, to represent processes not modeled by \despotic\ (e.g.~endothermic or exothermic chemical reactions). At the discretion of the user, \despotic\ can fix either $T_g$ or $T_d$ and solve for the other, or it can solve for both simultaneously. If either $T_g$ or $T_d$ is fixed, \despotic\ solves the equations using the secant method. If neither is fixed, it solves for $T_g$ and $T_d$ simultaneously using the \texttt{MINPACK} routine \texttt{hybrd1}, which implements the Powell hybrid method.

Finally, \despotic\ can compute the time-dependent thermal evolution of a cloud. Starting from an initial gas and dust temperature, \despotic\ can integrate equation (\ref{eq:tempevol}) for the gas temperature evolution. At the user's discretion, the calculation can be done either isochorically or isobarically. When evaluating the heating and cooling terms that appear on the right-hand side of equation (\ref{eq:tempevol}), \despotic\ assumes that both the level populations and the dust temperature reaches equilibrium instantaneously; the former are computed via the procedure described in Appendix \ref{sec:lines}, and the latter by the solution to equation (\ref{eq:dedtdust}) with $de_{d,\rm sp}/dt = 0$. \despotic\ also calculates the temperature-dependent specific heat $c_{v,\rm H}$ or $c_{p,\rm H}$ on the right-hand side using equation (\ref{eq:cv}).
It integrates equation (\ref{eq:tempevol}) using the \texttt{ODEPACK} routine \texttt{lsoda}, which automatically evaluates the stiffness of the system, and solves using a predictor-corrector method for non-stiff problems and backward differentiation formula methods for stiff problems.

\subsubsection{Chemical Evolution and Chemical Equilibria}

\despotic's implementation of chemistry has two parts. First, \despotic\ provides a series of routines that can integrate the chemical evolution equations (\ref{eq:chemeq}), either for a specified time interval or until the rates of change of all abundances are zero to within some specified tolerance. Second, \despotic\ provides a generic interface that can be used to implement arbitrary chemical networks. Once implemented, one can use the chemical evolution routines to integrate that network in an automated fashion. One basic network, that of \citet{nelson99a}, is included in the code repository.

\subsubsection{Line Profiles}

\despotic's final major capability is calculating line profiles for species in LTE. When performing this calculation, it accepts user-specified profiles for the number density of the emitting species, the bulk velocity, the non-thermal velocity dispersion, and the temperature as a function of radius. From these inputs, plus the identity of the line whose profile is to be computed, it calculates all the dimensionless quantities given in equations (\ref{eq:lineprofstart}) -- (\ref{eq:lineprofend}), and then numerically integrates equation (\ref{eq:transfer}) at a range of user-specified frequencies or velocities. The integration is performed via a call to the \texttt{ODEPACK} routine \texttt{lsoda}. \despotic\ then returns the CMB-subtracted intensity and brightness temperature as a function of frequency / velocity.

\subsection{Atomic and Molecular Data}

\despotic\ obtains the chemical data required for its computations (e.g., Einstein coefficients, reaction rate coefficients) from the Leiden Atomic and Molecular Database (LAMBDA; \citealt{schoier05a}). Access to the database is automated: \despotic\ automatically fetches whatever data files are needed without explicit user intervention. \despotic\ makes three approximations in situations where data from LAMDA is not available. First, for some species, LAMDA provides estimates only of collision rate coefficients for H$_2$, not for oH$_2$ and pH$_2$ separately, or it provides only oH$_2$ or pH$_2$. In such cases, \despotic\ assumes that the oH$_2$ and pH$_2$ collision rate coefficients are equal, and, if only generic H$_2$ rates are given, it sets both of them equal to those.

Second, for some species collision rate coefficients for H$_2$ are available, but collision rate coefficients for He are not. In this case \despotic\ assumes that He collision rate coefficients are related to those for H$_2$ by \citep{schoier05a}
\begin{equation}
k_{\rm He} = k_{\rm H_2} \left(\frac{\mu_{s-\rm H_2}}{\mu_{s-\rm He}}\right)^{1/2}
\end{equation}
where $\mu_{s-\rm H_2}$ is the reduced mass of the species $s$ with H$_2$, and similarly for $\mu_{s-\rm He}$.

Third, by default \despotic\ will not extrapolate collision rates outside the range of temperatures provided in the LAMDA tables. However, the user can override this default behavior, in which case \despotic\ will extrapolate by assuming that the downward collision rate coefficient varies as a powerlaw in the gas kinetic temperature. For linear molecules, a more accurate extrapolation motivated by a quantum mechanical treatment of the collision is possible (see the Section 6 of \citealt{schoier05a}), but no such treatment is available for non-linear molecules.

\section{Sample Applications}
\label{sec:examples}

In this section I provide some sample applications to demonstrate \despotic's capabilities. Each of these applications operates on one or more example clouds, whose properties are specified in Table \ref{tab:clouds}. The values given in this Table are intended to be examples only, but input files corresponding to each of them are included with the \despotic\ library to provide example templates that users can modify to set up their own clouds. The code to perform each of the example calculations listed below is also included with the \despotic\ download.

\begin{table*}
\caption{Sample clouds}
\label{tab:clouds}
\begin{tabular}{lcccc}
\hline
Cloud Name & MilkyWayGMC & ULIRG & ProtostellarCore & PostShockSlab \\
\hline
\multicolumn{4}{l}{Physical Properties} \\
$n_{\rm H}$ [cm$^{-3}$] & $10^2$ & $10^5$ & $10^2 - 10^8$ & $10^3$ \\
$N_{\rm H}$ [cm$^{-2}$] & $1.5\times 10^{22}$ & $10^{24}$ & $1.0\times 10^{23}$ & $1.5\times 10^{22}$ \\
$\sigma_{\rm NT}$ [km s$^{-1}$] & 2.0 & 80.0 & 0.1 & 0.5 \\
$T_g$ [K] & 8 & 45 & 8 & 250 \\
$T_d$ [K] & 8 & 60 & 8 & 8 \\
\\
\multicolumn{4}{l}{Composition} \\
$x_{\rm HI}$ & 0.0 & 0.0 & 0.0 & 0.0 \\
$x_{\rm oH2}$ & 0.1 & 0.1 & 0.1 & 0.1 \\
$x_{\rm pH2}$ & 0.4 & 0.4 & 0.4 & 0.4 \\
$x_{\rm He}$ & 0.1 & 0.1 & 0.1 & 0.1 \\
$x_e$ & 0.0 & 0.0 & 0.0 & 0.0 \\
$x_{\rm H^+}$ & 0.0 & 0.0 & 0.0 & 0.0 \\
\\
\multicolumn{4}{l}{Dust Properties} \\
$\alpha_{\rm GD}$ [erg cm$^3$ K$^{-3/2}$] & $3.2\times 10^{-34}$ & $3.2\times 10^{-34}$ &  $3.2\times 10^{-34}$ & $3.2\times 10^{-34}$\\
$\sigma_{d,10}$ [cm$^2$ H$^{-1}$] & $2.0\times 10^{-26}$ & $2.0\times 10^{-26}$ & $2.0\times 10^{-26}$ $2.0\times 10^{-26}$ \\
$\sigma_{d,\rm PE}$ [cm$^2$ H$^{-1}$] & $1.0\times 10^{-21}$ & $1.0\times 10^{-21}$ & $1.0\times 10^{-21}$ & $1.0\times 10^{-21}$ \\
$\sigma_{d,\rm ISRF}$ [cm$^2$ H$^{-1}$] & $3.0\times 10^{-22}$ & $3.0\times 10^{-22}$ & $3.0\times 10^{-22}$ & $3.0\times 10^{-22}$ \\
$Z'_d$ & 1.0 & 1.0 & 1.0 & 1.0 \\
$\beta_d$ & 2.0 & 2.0 & 2.0 & 2.0 \\
\\
\multicolumn{4}{l}{Radiation Field Properties} \\
$T_{\rm CMB}$ [K] & 2.73 & 2.73 & 2.73 & 2.73 \\
$T_{\rm rad,dust}$ [K] & 0.0 & 60.0 & 8.0 & 8.0 \\
$\zeta$ [s$^{-1}$ H$^{-1}$] & $1.0\times 10^{-16}$ & $2.0\times 10^{-15}$ & $2.0\times 10^{-17}$ & $2.0\times 10^{-17}$ \\
$\chi$ & 1.0 & $1.0\times 10^4$ & 1.0 & 1.0 \\
\\
\multicolumn{4}{l}{Emitting Species Abundances} \\
CO & $1.0\times 10^{-4}$ & $1.0\times 10^{-4}$ & $1.0\times 10^{-4}$ & $1.0\times 10^{-4}$ \\
$^{13}$CO & $5.0\times 10^{-7}$ & $5.0\times 10^{-7}$ & $5.0\times 10^{-7}$ & $5.0\times 10^{-7}$ \\
C$^{18}$O & - & - & $5.0\times 10^{-8}$ & $5.0\times 10^{-8\,*}$ \\
C & - & - & $5.0\times 10^{-7}$ & $5.0\times 10^{-7\,*}$ \\
O & - & - & $5.0\times 10^{-6}$ & $5.0\times 10^{-6}$\\
CS & - & - & $1.0\times 10^{-8}$ & $1.0\times 10^{-8\, *}$ \\
HCO$^+$ & - & - & $1.0\times 10^{-8}$ & $1.0\times 10^{-8\, *}$ \\
pNH$_3$ & - & - & $1.0\times 10^{-8}$ & $1.0\times 10^{-8\, *}$ \\
oNH$_3$ & - & - & $1.0\times 10^{-8}$ & $1.0\times 10^{-8\, *}$ \\
pH$_2$CO & - & - & $1.0\times 10^{-8}$ & $1.0\times 10^{-8\, *}$ \\
oH$_2$CO & - & - & $1.0\times 10^{-8}$ & $1.0\times 10^{-8\, *}$ \\
pH$_2$O & - & - & $1.0\times 10^{-8}$ & $1.0\times 10^{-8\, *}$ \\
oH$_2$O & - & - & $1.0\times 10^{-8}$ & $1.0\times 10^{-8\, *}$ \\
\hline
\end{tabular}

\medskip

The table gives initial properties for the example cloud models used in \S~\ref{sec:examples}. For applications where $T_g$ and $T_d$ are fixed, the values given in the table are the values used; for applications where $T_g$ and $T_d$ are to be calculated, they are used as initial guesses. For the ProtostellarCore model, the density is given as a range because a range of models are run. The abundances in this model have been chosen to roughly match those recommended in \citet{goldsmith01a}. For the PostShockSlab model, emitting species marked with asterisks indicate species from which line emission is computed, but that are ignored for the purposes of calculating the thermal evolution. The molecular data from LAMDA used in evaluating these models are taken from the following sources: CO, $^{13}$CO, and C$^{18}$O: \citet{yang10a}; C: \citet{schroder91a} and \citet{staemmler91a}; O: \citet{jaquet92a}; CS: \citet{turner92a}; HCO$^+$: \citet{flower99a}; NH$_3$: \citet{danby88a}; H$_2$CO: \citet{green91a}; H$_2$O: \citet{daniel11a}. 
\end{table*}

\subsection{CO Spectral Line Energy Distributions}

\begin{figure}
\includegraphics[width=84mm]{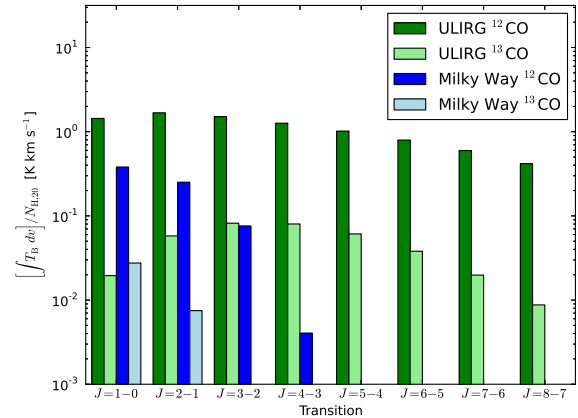}
\caption{
\label{fig:coSLED}
Spectral line energy distribution for the first 8 rotational transitions of CO and $^{13}$CO, computed for the models MilkyWayGMC and ULIRG described in Table \ref{tab:clouds}. The plot shows the velocity-integrated brightness temperature in each line normalized by $N_{\rm H,20} = N_{\rm H}/10^{20}$ cm$^{-2}$. The contribution of the CMB has been subtracted off.
}
\end{figure}

As a first example of \despotic's capabilities, Figure \ref{fig:coSLED} shows a calculation of CO and $^{13}$CO spectral line energy distributions (SLEDs) for the MilkyWayGMC and ULIRG clouds described in Table \ref{tab:clouds}. For this computation, the gas temperature is left fixed to the input value, and the level populations are computing using the escape probability formalism. As expected, all lines of the ULIRG are much brighter due to its higher gas kinetic temperature and velocity dispersion -- to first order, the velocity-integrated brightness temperature of an optically thick line is simply the product of those two. In addition, the falloff in luminosity with $J$ is much slower for the ULIRG than for the Milky Way cloud. This is as a result of the much higher density and temperature of the ULIRG. The former allows its higher levels to be close to thermally populated, and the latter causes their thermal populations to be large. We also see that the $^{12}$CO(1-0) to $^{13}$CO(1-0) ratio is larger for the ULIRG than for the Milky Way model, reflecting the higher optical depth of the ULIRG. At higher $J$, where the optical depth drops, the line ratios of the two isotopomers vary less between the two models.

Note that this computation for $^{12}$CO(1-0) is equivalent to calculating the CO ``X-factor" that relates CO intensities to cloud masses and column densities. The values calculated by \despotic\ are $X_{\rm CO} = 2.6\times 10^{20}$ cm$^{-2} / \left(\mbox{K km s}^{-1}\right)$ for MilkyWayGMC, and  $X_{\rm CO} = 7.0\times 10^{19}$ cm$^{-2} / \left(\mbox{K km s}^{-1}\right)$. This is in line with other theoretical and observational estimates for normal galaxies and ULIRGs, respectively \citep{bolatto13a}. One should be wary of reading too much into this result, since neither the chemical and thermal states of the clouds have been specified by hand. This computation should be done by combining a three-dimensional simulation with chemical post-processing to determine the chemical state of the clouds self-consistently, and then using \despotic\ or a similar package to calculate the resulting gas temperature and line radiation \citep[e.g.][]{narayanan11a, narayanan12a, shetty11a, shetty11b, feldmann12a, feldmann12b}.

\subsection{Temperatures of Protostellar Cores}
\label{sec:core}

\begin{figure}
\begin{center}
\includegraphics[width=84mm]{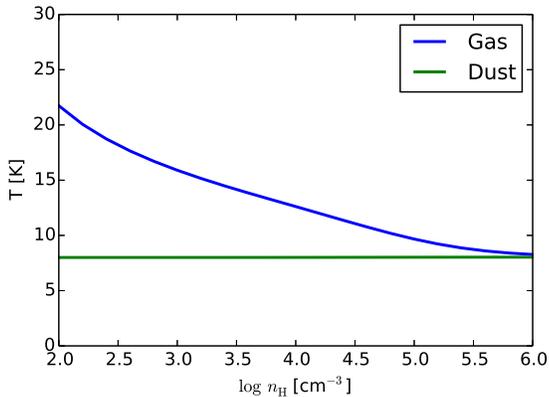}
\end{center}
\caption{
\label{fig:coreTemp}
Equilibrium gas and dust temperatures versus density for the ProtostellarCore model described in \S~\ref{sec:core}.
}
\end{figure}

\begin{figure*}
\includegraphics[width=4in]{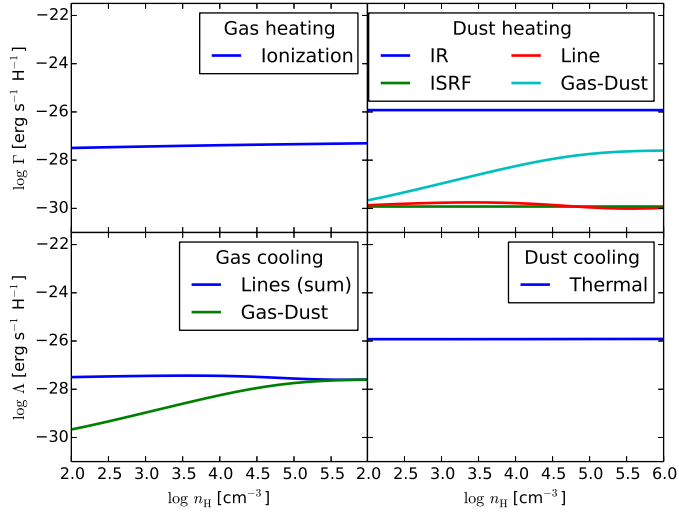}
\caption{
\label{fig:coreHeatCool}
Values of the various heating and cooling terms for dust and gas in the ProtostellarCore models, calculated at the equilibrium temperatures shown in Figure \ref{fig:coreTemp}. The panels show gas heating terms, gas cooling terms, dust heating terms, and dust cooling terms, as indicated in the legends. The terms shown are gas ionization heating, $\Gamma_{\rm ion}$, dust IR heating, $\Gamma_{d,\rm IR}$, dust ISRF heating $\Gamma_{d,\rm ISRF}$, dust line heating $\Gamma_{d,\rm line}$, gas line cooling, $\Lambda_{\rm line}$, dust thermal cooling, $\Lambda_d$, and dust-gas energy exchange, $\Psi_{\rm gd}$. The calculations also include gravitational and photoelectric heating, but these terms are below the plotted range.
}
\end{figure*}

\begin{figure}
\includegraphics[width=84mm]{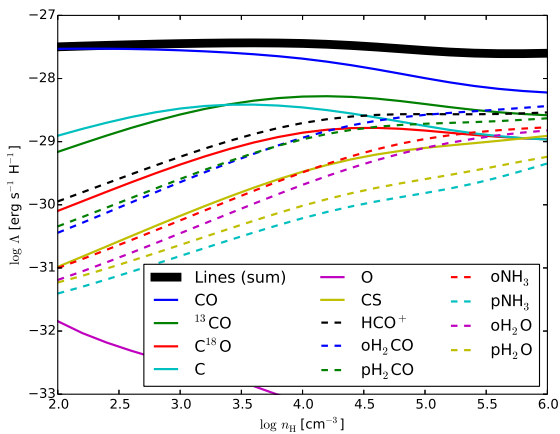}
\caption{
\label{fig:coreLines}
Contributions to the overall line cooling rate for individual atomic molecular species in the ProtostellarCore model. All line cooling rates are computed at the equilibrium temperatures shown in Figure \ref{fig:coreTemp}. Note that these rates are computed for constant abundances, and thus do not properly account for depletion at high densities. They are therefore likely to be overestimates at the high-density end, as discussed in \S~\ref{sec:core}.
}
\end{figure}

As a second example application, I use \despotic\ to calculate the equilibrium gas and dust temperatures in protostellar cores as a function of density, using the algorithms outlines in \S~\ref{sec:therm}. In this calculation I include a large number of cooling species (see Table \ref{tab:clouds}) in order to assess their density- and temperature-dependent contribution to cores' thermal balance. For this calculation I use the ProtostellarCore model in Table \ref{tab:clouds}. I then compute a grid of models with densities in the range $n_{\rm H} = 10^2 - 10^6$ cm$^{-3}$ in steps of 0.2 dex. For each model, I compute the equilibrium gas and dust temperatures, and, once the equilibrium has been calculated, I record the values of all the heating and cooling terms.

Figure \ref{fig:coreTemp} shows the equilibrium temperatures as a function of density, Figure \ref{fig:coreHeatCool} shows the contributions of the various heating and cooling processes, and Figure \ref{fig:coreLines} further subdivides the line cooling into the contributions made by individual species. The plots illustrate a number of phenomena. First, the gas temperature is relatively high at low densities, and drops as the density increases. At densities below $\sim 10^4$ cm$^{-3}$ this drop is driven by increasingly effective line cooling. Between $10^4$ and $10^5$ cm$^{-3}$, dust-gas collisions become competitive with line cooling, and lock the dust and gas temperature together, such that dust-gas energy exchange becomes dominant in setting the temperature. The dust in turn is always locked close to the infrared radiation field temperature, because the IR heating rate and thermal cooling rate both exceed all other sources and sinks of energy for the dust by orders of magnitude.

In terms of molecular line cooling, at low densities the dominant coolants are CO, $^{13}$CO, and C. As the density rises and the dust temperature drops, these become less important because dust coupling lowers the gas temperature. This makes it more difficult to excite the higher $J$ lines that have lower optical depths. At the same time, other species make an increasing contribution to the cooling as the density approaches their critical densities and begins to provide efficient collisional excitation. However, this example also illustrates one of \despotic's limitations. These calculations assume density- and temperature-independent abundances, and do not properly model the effects of freeze-out onto grain surfaces. Over the density range I have explored freeze-out is probably significant only for CS, since sulfur-bearing molecules begin to freeze out at densities above $\sim 10^3-10^4$ cm$^{-3}$, but carbon- and nitrogen-bearing ones do not experience significant freeze-out until the density rises above $\sim 10^6$ cm$^{-3}$ and $\sim 10^7-10^8$ cm$^{-3}$, respectively \citep[e.g.][]{bergin97a}. Once could include freeze-out effects by defining an appropriate chemistry network, but the simple \citet{nelson99a} network that \despotic\ currently implements does not model these effects.

This is the most computationally-intensive of the example applications provided, due to the high optical depth and the large number of molecular coolants included. The majority of the computational effort involves iterating to obtain the level populations at high optical depth. Evaluating the entire grid of 21 models requires a bit under 5 minutes. However, since only a few chemical species are actually important to the thermal balance, one could obtain the results far more quickly simply by ignoring the large number of energetically-unimportant species when calculating the temperature, and only calculating their line luminosities once the temperatures have converged. \despotic\ includes a capability to mark certain species as energetically-unimportant, allowing them to be treated in precisely this manner, and I demonstrate this capability in the next example.

\subsection{Time-Dependent Cooling of Post-Shock Gas}
\label{sec:shockCool}

A third example, which makes use of \despotic's ability to calculate time-dependent temperature evolution (\S~\ref{sec:therm}), is to calculate the cooling of out-of-equilibrium gas. I consider a slab of gas whose properties are given by the PostShockSlab model in Table \ref{tab:clouds}. At time $t=0$, the gas has just been shock-heated to an out-of-equilibrium temperature of $250$ K, and I calculate the time evolution of its temperature and line emission thereafter, assuming that the gas is isobaric and using a slab geometry to compute escape probabilities. In calculating the thermal evolution I include only the energetically-dominant coolants CO, $^{13}$CO, and O, but I also periodically compute the line emission of a large number of other species as well (see Table \ref{tab:clouds} for the full list). By making this assumption, the total computer time required to evolve the model 40 kyr, including periodic calculation of emission from many lines, is $\sim 10$ minutes.

\begin{figure}
\includegraphics[width=84mm]{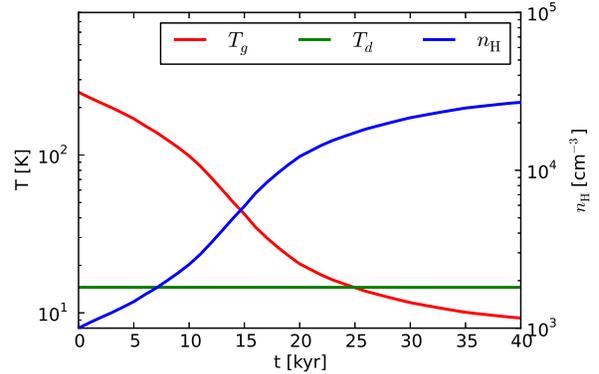}
\caption{
\label{fig:shockCool_temp}
Gas temperature, dust temperature, and gas density versus time for isobaric cooling of the PostShockSlab model, as described in \S~\ref{sec:shockCool}.
}
\end{figure}

\begin{figure}
\includegraphics[width=84mm]{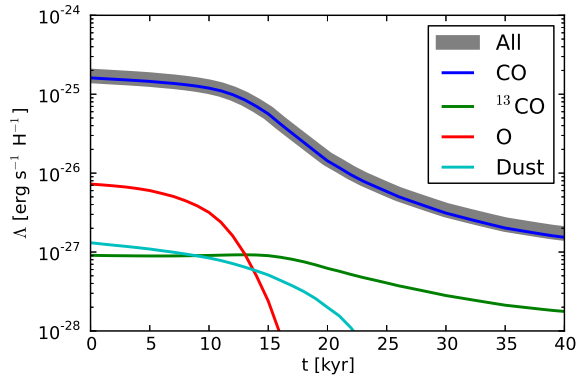}
\caption{
\label{fig:shockCool_coolants}
Rates of cooling provided by CO lines, $^{13}$CO lines, O lines, and dust versus time, for the PostShockSlab model shown in Figure \ref{fig:shockCool_temp}. The gray thick line shows the sum of all coolants, including all a number of lines that are not shown because they all below the range of cooling rates plotted.
}
\end{figure}

\begin{figure}
\includegraphics[width=84mm]{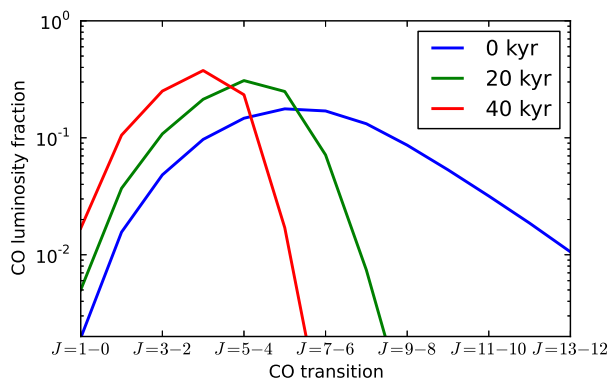}
\caption{
\label{fig:shockCool_coSLED}
CO line spectral line energy distribution for the PostShockSlab model shown in Figure \ref{fig:shockCool_temp}. Each of the lines shows the relative contributions of the indicated rotational transitions of CO to the total cooling rate at the indicated times of 0 kyr, 20 kyr, and 40 kyr. Contributions are normalized so that the sum over all transitions is unity.
}
\end{figure}

Figure \ref{fig:shockCool_temp} shows the gas temperature, dust temperature, and gas density versus time as computed by \despotic\ for this initial condition. Figure \ref{fig:shockCool_coolants} shows the contributions of various species to the cooling. As the plot shows, cooling is dominated by CO lines, with minor contributions from $^{13}$CO, O, and dust, and negligible contributions from all other sources. In Figure \ref{fig:shockCool_coSLED} I further examine the cooling, by showing how the CO spectral line energy distribution changes with time. As the plot shows, the SLED initially peaks near $J = 7-6$, and moves to a cooler SLED at time passes. At the final time shown, $J=3-2$ is the dominant coolant. Note that this differs from the result shown in Figure \ref{fig:coSLED} for a typical GMC because the post-shock slab we are considering has a significantly lower velocity dispersion and a significantly higher density. Both of these favor cooling through higher $J$ lines, the former because it increases the optical depth for low $J$ lines, and the latter because it helps to thermalize higher $J$ states.

\subsection{Carbon-Oxygen Chemistry}
\label{sec:gmcchem}

\begin{figure}
\includegraphics[width=84mm]{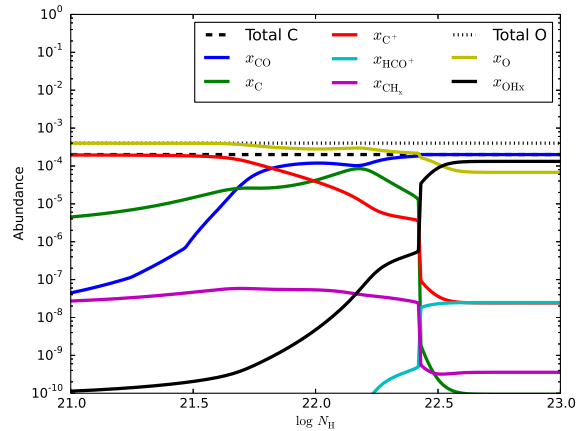}
\caption{
\label{fig:gmcchem}
Abundances of various carbon- and oxygen-bearing species as a function of gas column column density, computed using \despotic's chemistry module following the procedure described in Section \ref{sec:gmcchem}.
}
\end{figure}

The next sample application demonstrates \despotic's chemistry capability. For this test, I use the physical parameters for the MilkyWayGMC model listed in Table \ref{tab:clouds}, except that I reduce the cosmic ray ionization rate to $\zeta=3\times 10^{-17}$ s$^{-1}$ and raise the gas temperature to 10 K. I then consider a range of column densities $N_{\rm H} = 10^{21} - 10^{23}$ cm$^{-2}$, in steps of $0.01$ dex, and use the \citet{nelson99a} chemical network implemented in \despotic\ to calculate the equilibrium chemical state of the cloud. The column density range is chosen to model the transition from a composition dominated by H$_2$, C$^+$, and O (the so-called ``dark gas" -- \citet{wolfire10a}) to one dominated by H$_2$ and CO, as found in the interiors of molecular clouds. The total execution time of the calculation, which involves finding the equilibrium chemical state 201 times, was roughly a minute.

Figure \ref{fig:gmcchem} shows the result. As shown in the figure, at low column density the chemical state is such that the carbon is mostly C$^+$ and the oxygen is mostly O. As the column density increases, the carbon shifts into C and CO as the dominant states, while the oxygen also shifts into CO and OH$_x$. The chemical transition is driven by the decreasing rate of photodissociation as the column density increases, which, following \citeauthor{nelson99a}'s prescription, is handled using the tabulated shielding functions of \citet{van-dishoeck88a}. This test demonstrates \despotic's ability to perform limited astrochemistry calculations.

\subsection{Inverse P Cygni Profiles}
\label{sec:pcygni}

As a final application, I use \despotic\ to calculate line profiles in a collapsing protostellar core. For this example, I consider a core with a radius of $R=0.02$ pc with a velocity profile $v(r) = -0.4 (r/R) \hat{r}$ km s$^{-1}$. The temperature profile is $T(r) = 8+12\exp(-2r^2/R^2)$ K, so that the temperature reaches a peak of 20 K at the center, dropping close to 8 K at large radii. The core also has a position-independent non-thermal velocity dispersion of $0.2$ km s$^{-1}$. I use a uniform density $n_{\rm H} = 3\times 10^6$ cm$^{-3}$.

For this core I use \despotic\ to compute the profiles of the HCN(1-0) and N$_2$H$^{+}$ lines. This combination of lines is often used to measure infall motions \citep[e.g.,][]{sohn07a}, as their overall spatial distributions in a protostellar core are thought to be quite similar on chemical grounds, but the HCN(1-0) tends to be marginally optically thick and develop inverse P Cygni profiles, while the N$_2$H$^+$ tends to be optically thin and show symmetric profiles. I adopt abundances $x_{\rm HCN} = 2\times 10^{-9}$ and $x_{\rm N_2H^+} = 2\times 10^{-9}$ based on the models of \citet{lee04a}. 

\begin{figure}
\includegraphics[width=84mm]{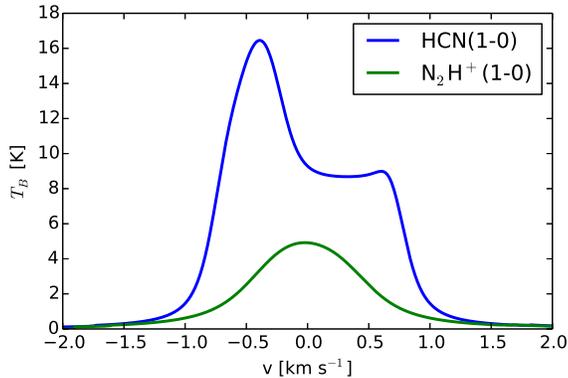}
\caption{
\label{fig:pcygni}
Brightness temperature versus velocity relative to line center for the lines HCN($1-0$) and N$_2$H$^+$($1-0$) produced by a collapsing protostellar core. The contribution of the CMB has been subtracted off. Details of the core parameters are given in \S~\ref{sec:pcygni}.
}
\end{figure}

Figure \ref{fig:pcygni} show the line profiles computed by \despotic. As expected, the marginally optically thick HCN line produces a double-peaked asymmetric inverse P Cygni profile, indicative of infall. The N$_2$H$^+$ line is optically thin and produces a symmetric profile of lower total intensity. The total time required to perform the computation is $\sim 10$ s.

\section{Limitations and Caveats}
\label{sec:limitations}

While \despotic\ provides reasonable estimates of the thermal behavior and spectra of interstellar clouds over a wide range of environments, it also has significant limitations, which I discuss here as a warning to potential users. The major limitations of the code are:
\begin{itemize}
\item \despotic's treatment of dust temperatures is very crude in the regime of clouds that are optically thick to their own cooling radiation. In such clouds the dust temperature will be determined largely by the value of $T_{\rm rad,dust}$ that the user selects. If a user requires accurate dust temperatures in such clouds, he or she is advised to use a code like \texttt{dusty} \citep{ivezic97a} to calculate the dust temperature and radiation field within the cloud, then use this to set $T_{\rm rad,dust}$ for the purposes of a \despotic\ calculation.

\item \despotic\ neglects the contribution of the dust radiation field to the photon occupation number when calculating level populations, on the grounds that, because dust optical depths are small at low frequencies, such fields are often highly sub-thermal at the low frequencies where most important molecular lines lie. However, in some circumstances, e.g.~protostellar disks \citep{krumholz07d}, the column density is so high that dust optical depths can exceed unity even at frequencies as low as $20$ GHz. In such environments excitation and de-excitation of molecules by interaction with the infrared field is non-negligible, and \despotic\ will not give accurate results.

\item \despotic\ uses a one-zone model, and this is not capable of capturing effects that depend on radiative transfer. In particular, \despotic\ cannot handle maser emission, and it cannot handle effects on the line shape that arise from spatially-variable departures from LTE.

\item The repository version of \despotic\ includes only a single, very simple chemical network. This can be used to make reasonable predictions for carbon and oxygen chemistry in H$_2$-dominated environments, but not for other species or in other environments. It is up to the user to either input  chemical abundances directly, or to implement chemical networks appropriate for the environment he or she wishes to simulate. The results \despotic\ produces will only be as good as those abundances or networks. More subtly, \despotic\ does include the effects of selective chemical destruction of excited states on line emission, and it does not include any heating or cooling of the gas or dust as a result of chemical reactions, such as heating of dust grains by exothermic formation of H$_2$ on grain surfaces \citep[e.g.][]{lesaffre05a}. \despotic\ provides a mechanism to include chemical heating and cooling, since the user can specify arbitrary additional heating and cooling terms, but it is up to the user to determine whether there are any energetically-important chemical reactions for the problem under consideration, and, if so, to implement the necessary code.
\end{itemize}

\section{Summary}
\label{sec:summary}

I introduce \despotic, a Python-based, open-source software library for calculating spectra, heating and cooling rates, and time-dependent and time-independent thermal properties of optically thick interstellar clouds. \despotic\ includes all the dominant heating and cooling processes for both gas and dust over a wide range of interstellar environments, and can be used to conduct both fast sweeps of parameter space and interactive explorations within an interactive Python environment. It is intended to allow theoretical investigators to obtain approximate values of parameters such as cloud temperatures, major heating and cooling processes, and observable line emission, without the difficulty and time investment of developing their own statistical and thermal equilibrium codes, and with significantly less investment of CPU and human time than would be required to approach such problems using a detailed PDR code. \despotic\ is under continued development, and additional features capabilities will be released to the community as they are implemented.

\section*{Acknowledgements}

I thank the creators and maintainers of the Leiden Atomic and Molecular Database, F.~Sch\"oier, F.~van der Tak, E.~van Dishoeck, and J.~Black, for providing that valuable resource. I thank B.~Draine for helpful suggestions regarding modeling of dust, and F.~van der Tak for helpful suggestions on the manuscript and advice on \texttt{RADEX}. I acknowledge support from the Alfred P.~Sloan Foundation,  the NSF through CAREER grant AST-0955300, and NASA through Astrophysics Theory and Fundamental Physics Grant NNX09AK31G.

\bibliographystyle{mn2e}
\bibliography{refs}

\appendix

\section{Specific Heats}
\label{sec:cv}

Calculating the time evolution of the temperature requires knowing the specific heat per H nucleus at constant volume $c_{v,\rm H}$, defined by
\begin{equation}
c_{v,\rm H} = \frac{1}{n_{\rm H}} \left(\frac{\partial e_g}{\partial T}\right)_\rho,
\end{equation}
where $e_g$ is the gas internal energy per unit volume, given by
\begin{equation}
e_g = \sum_s n_s k_B T \frac{d\ln z_s}{d\ln T},
\end{equation}
where the sum runs over species $s$, $n_s$ is the number density of species $s$, and $z_s$ is the partition function per unit volume for that species. The latter is given by
\begin{equation}
z_s = Z_{s, \rm trans}  Z_{s,\rm rot}  Z_{s,\rm vib}  Z_{s,\rm spin},
\end{equation}
where the terms appearing in the equation above are the partition functions for the translation, rotational, vibrational, and spin degrees of freedom of species $i$. In principle we should also include a term describing electronic degrees of freedom, but at the relatively low temperatures for which \despotic\ is intended, we can safely assume that these are not excited. For all the species included in \despotic\ except molecular hydrogen (i.e.~for H~\textsc{i}, He, H$^+$, and $e$), the contribution of the specific heat is trivial, because all of the partition functions except translation and spin are unity, and the spin term is temperature-independent. Thus for all these species
\begin{equation}
\frac{\partial \ln z_s}{\partial \ln T} = \frac{\partial \ln Z_{s,\rm trans}}{\partial \ln T} = \frac{3}{2}.
\end{equation}
For ortho- and para-H$_2$ on the other hand, $Z_{\rm rot}$ and $Z_{\rm vib}$ are not unity \citep{black75a, boley07a, tomida13a}:
\begin{eqnarray}
Z_{\rm oH_2, rot} & = & \sum_{J\,\rm odd} 3(2J+1) \exp\left[-\frac{J(J+1)\theta_{\rm rot}}{T}\right] \\
Z_{\rm pH_2, rot} & = & \sum_{J\,\rm even} (2J+1) \exp\left[-\frac{J(J+1)\theta_{\rm rot}}{T}\right] \\
Z_{\rm H_2, vib} & = & \frac{1}{1-\exp(-\theta_{\rm vib}/T)}
\end{eqnarray}
where $\theta_{\rm rot} = 85.3$ K and $\theta_{\rm vib} = 5984$ K. Note that the vibrational partition function is the same for ortho- and para-H$_2$, but the rotational partition functions are different. With these partition functions, the energy per unit volume including all species is
\begin{eqnarray}
\frac{e_g}{k_B} & = & \frac{3}{2} T \sum_s n_s  + n_{\rm pH_2} \left(\frac{T^2}{Z_{\rm pH_2}}\frac{\partial Z_{\rm pH_2,rot}}{\partial T}\right)
\nonumber
\\
& & {} +
n_{\rm oH_2} \left(\frac{T^2}{Z_{\rm oH_2}}\frac{\partial Z_{\rm oH_2,rot}}{\partial T}\right) 
\nonumber
\\
& & {}
+
\left(n_{\rm pH_2}+n_{\rm oH_2}\right) \theta_{\rm vib} \frac{\exp(-\theta_{\rm vib}/T)}{1-\exp(-\theta_{\rm vib}/T)},
\end{eqnarray}
where again the sum runs over all all species.

Deriving the specific heat $c_v$ from this expression requires making an assumption about how the number densities of ortho- and para-H$_2$ vary with temperature. At the low temperatures found in interstellar clouds, there is generally no efficient mechanism for converting between the two states, and thus the most reasonable assumption is that these number densities are temperature-independent. Observations showing that the ortho- to para- ratio in molecular clouds is far from equilibrium \citep[e.g.][]{neufeld06a, pagani11a, dislaire12a} support this assumption. For temperature-independent values of $n_{\rm pH_2}$ and $n_{\rm oH_2}$, we therefore have
\begin{eqnarray}
\frac{c_{v,\rm H}}{k_B} & = & \frac{3}{2}\sum_s x_s + x_{\rm pH_2} \frac{\partial}{\partial T}\left(\frac{T^2}{Z_{\rm pH_2}}\frac{\partial Z_{\rm pH2}}{\partial T}\right)
\nonumber \\
& & {}
+ x_{\rm oH_2} \frac{\partial}{\partial T}\left(\frac{T^2}{Z_{\rm oH_2}}\frac{\partial Z_{\rm oH2}}{\partial T}\right)
\nonumber \\
& & {}
+ \left(x_{\rm pH_2}+x_{\rm oH_2}\right) \frac{\theta_{\rm vib}^2\exp(-\theta_{\rm vib}/T)}{T^2[1-\exp(-\theta_{\rm vib}/T)]^2}.
\label{eq:cv}
\end{eqnarray}
Note that this expression involves the abundances ratios $x$ rather than number densities $n$ because we have normalized all quantities to the number density of H nuclei. The specific heat at constant pressure is simply $c_{p,\rm H}/k_B = c_{v,\rm H}/k_B + 1$.

\section{Heating and Cooling Processes}
\label{sec:heatcool}

Here I give explicit formulae for all the heating processes included in \despotic. In the following description, all heating, cooling, and energy exchange rates are given as energies per H nucleus per unit time.

\subsection{Gas}
\label{sec:gasheatcool}

\subsubsection{Ionization Heating}

Gas can gain energy through ionization heating; in this process primary electrons with energies produced when the gas is ionized by cosmic rays or hard x-rays thermalize, adding energy. The rate at which this process adds energy is given by
\begin{equation}
\Gamma_{\rm ion} = \zeta q_{\rm ion},
\end{equation}
where $q_{\rm ion}$ is the energy added per primary ionization. The value of $q_{\rm ion}$ in turn depends on the bulk chemical composition of the gas, which determines how much of a primary electron's $\approx 37$ eV of energy is lost via radiation rather than transformed into heat. This problem has been discussed by a number of authors \citep{dalgarno72a, glassgold73a, wolfire95a, dalgarno99a, wolfire10a, glassgold12a}. In predominantly atomic regions, the main pathway to thermalization is Coulomb scattering of the primary electron off other free electrons, and collisional excitation of H and He by the primary electron followed by collisional de-excitation of the excited atom. In this regime \despotic\ uses the approximation recommended by \citet{draine11a},
\begin{equation}
q_{\rm ion,HI} \approx 6.5\mbox{ eV} + 26.4\mbox{ eV}\left(\frac{x_e}{x_e+0.07}\right)^{1/2}.
\end{equation}

In molecular regions the situation is far more complicated due to the additional thermalization channels provided by excitation of the rotational and vibrational levels of H$_2$ (followed by collisional de-excitation), by dissociation of H$_2$, and by chemical heating, in which primary electrons produce reactive ions such as H$_2^+$, H$^+$, and He$^+$ that subsequently undergo exothermic reactions with neutrals such as CO, H$_2$O, and O. In this case $q_{\rm ion}$ becomes a complex function of the gas density and temperature, and the abundances of various species, and ranges from $\sim 10 - 20$ eV as these quantities change \citep{glassgold12a}. Given the complexity of the problem, and the level of inaccuracy inherent in any one-zone model, \despotic\ relies on a simple piecewise fit to the numerical results of \citet[their Table 6]{glassgold12a} on the density-dependence of $q_{\rm ion}$ in molecular regions:
\begin{equation}
\frac{q_{\rm ion,H_2}}{\mbox{eV}} \approx 
\left\{
\begin{array}{ll}
10, & \log n_{\rm H} \leq 2 \\
10 + 3 (\log n_{\rm H} - 2)/2, & 2 \leq \log n_{\rm H} < 4 \\
13 + 4 (\log n_{\rm H} - 4)/3, & 4 \leq \log n_{\rm H} < 7 \\
17 + 1(\log n_{\rm H} - 7)/3, & 7 \leq \log n_{\rm H} < 10\\
18, & \log n_{\rm H} \geq 10
\end{array}
\right.,
\end{equation}
where the values of $n_{\rm H}$ in the above expression are in units of cm$^{-3}$.

To handle the case where the composition includes both molecular and atomic gas, \despotic\ assumes that the atomic and molecular regions are physically separated (which, depending on the physical situation, may or may not be a good assumption). In this case the total heating rate can be computed simply by summing the heating rates in the atomic- and molecular-dominated regions, weighted by their number fractions:
\begin{equation}
q_{\rm ion} = x_{\rm HI} q_{\rm ion,HI} + 2 (x_{\rm oH_2} + x_{\rm pH_2}) q_{\rm ion,H_2}.
\end{equation}

\subsubsection{Photoelectric Heating}

Gas can also gain energy through grain photoelectric heating, whereby a primary electron ejected from a dust grain by a far-ultraviolet (FUV) photon thermalizes with the gas. Unlike cosmic rays, the FUV photons responsible for photoelectric heating can be attenuated by dust rather easily, and the photoelectric heating rate therefore depends on four factors: the strength of the ISRF, the abundance of dust grains, the amount of dust shielding, and the energy yield per photoelectron; as with cosmic ray heating, the latter value has been estimated by numerous authors \citep{watson72a, de-jong77a, tielens85a, bakes94a, wolfire03a}. To account for dust shielding, which obviously varies from point to point within a real cloud, \despotic\ uses the simple approximation proposed by \citet{krumholz11b}, whereby the $8-13.6$ eV photons responsible for photoelectron production are considered to be attenuated by half the mean extinction of the cloud. Since the dust opacity is relatively flat across this energy range ($\sim 50\%$ variation in the models of \citealt{draine03b}), we can assign a single cross section $\sigma_{d,\rm PE}$, which is $\sim 10^{-21}$ cm$^2$ H$^{-1}$ for Milky Way dust. This value is near the middle of the range found in the models of \citet{draine03b}. With this approximation, the photoelectric heating rate becomes
\begin{equation}
\Gamma_{\rm PE} = 4.0\times 10^{-26} \chi Z'_d e^{-(1/2)N_{\rm H}\sigma_{d,\rm PE}} \mbox{ erg s}^{-1}\mbox{ H}^{-1}.
\end{equation}

\subsubsection{Gravitational Heating}

A third possible source of heating is adiabatic compression. This obviously depends on the hydrodynamics of the flow, something that is not naturally included in a one-zone model like that used in \despotic. However, this effect is calculable in the special case of compression due to gravitational contraction, as in protostellar cores for example. In this case the heating rate may be computed using the approximation introduced by \citet{masunaga98a},
\begin{equation}
\Gamma_{\rm grav} = C_1 c_s^2  \muH \mH \sqrt{4\pi G\rho},
\end{equation}
where $C_1$ is a dimensionless constant of order unity that depends on the nature of the gravitational collapse. From their numerical calculations, \citet{masunaga98a} find $C_1 \approx 1.0$. Since in general most interstellar clouds are not in a state of collapse, by default \despotic\ does not include gravitational contraction heating, and sets $C_1 = 0$. However, users do have the option of overriding this default.

\subsubsection{Line Cooling}

The primary cooling mechanism for gas is line radiation. For each emitter species $s$, there is a rate of line cooling $\Lambda_s$, so that the total line cooling rate is
\begin{equation}
\Lambda_{\rm line} = \sum_s \Lambda_s.
\end{equation}
I defer a calculation of $\Lambda_s$ to \S~\ref{sec:lines}.

\subsubsection{Dust-Gas Energy Exchange}

Finally, gas can either heat or cool by exchanging energy with the dust via collisions. The gas-dust energy exchange rate is given by
\begin{equation}
\Psi_{\rm gd} = \alpha_{\rm gd} f_{\rm cl} n_{\rm H} T_g^{1/2} (T_d - T_g),
\end{equation}
where $\alpha_{\rm gd}$ is the grain-gas coupling coefficient and the sign convention is that positive values correspond to heating of the gas and cooling of the dust. Note the presence of the clumping factor $f_{\rm cl}$, since this is a collisional process. The coupling constant depends on the grain abundance, chemical composition, size distribution, and charge state. For Milky Way dust, \citet{goldsmith01a} recommends a value $\alpha_{\rm gd} = 3.2\times 10^{-34}$ erg cm$^3$ K$^{-3/2}$ for H$_2$-dominated regions, and \citet{krumholz11b} estimate a value of $1.0\times 10^{-33}$ erg cm$^3$ K$^{-3/2}$ for H~\textsc{i}-dominated ones, with the difference arising due to the change in both the number and mean mass of free particles between H~\textsc{i} and H$_2$-dominated regions.

\subsection{Dust}

\subsubsection{Cooling by Thermal Radiation}

Dust grains can lose energy via thermal continuum radiation. To compute the cooling rate, consider a population of spherical grains with distribution of radii $a_g$ given by $dn/da_g$, where we normalize the distribution function such that $n_d = \int (dn/da_g)\, da_g$ is the total number density of dust grains. Let $Q_\nu(a_g)$ be the absorption efficiency for absorption of radiation of frequency $\nu$, so that the cross section of the grain to radiation of frequency $\nu$ is $\sigma_\nu(a_g) = \pi a_g^2 Q(\nu)$. Further let $\langle Q(a_g)\rangle_T = \int B_\nu(T) Q_\nu(a_g) \, d\nu / \int B_\nu(T) \, d\nu$ be the Planck-weighted mean efficiency, where $B_\nu(T)$ is the Planck function evaluated at temperature $T$. Given this definition, we can write the rate of thermal radiation cooling from dust grains of temperature $T_d$ as
\begin{eqnarray}
\Lambda_{d, \rm thin} & = & \left[\frac{1}{n_{\rm H}} \int \frac{dn}{da_g} \langle Q_\nu(a_g)\rangle_T \pi a_g^2 \, da_g\right] c a T_d^4 \\
& \equiv & \sigma_d(T_d) c a T_d^4,
\end{eqnarray}
we have defined the term in square brackets to be the mean dust cross section per H nucleus $\sigma_d(T_d)$. This expression assumes that the cloud is optically thin to its own cooling radiation; we treat the optically thick regime below. We approximate that $\sigma_d(T_d)$ will vary as a powerlaw with $T_d$, and we therefore write
\begin{equation}
\sigma_d(T_d) = \sigma_{d,10} \left(\frac{T_d}{10\mbox{ K}}\right)^{\beta}.
\end{equation}
For Milky Way dust, typical opacities are $\sigma_{d,10}\approx 2\times 10^{-25}$ cm$^2$ H$^{-1}$ \citep{pollack94a, semenov03a}, and for temperatures $T_d$ such at $hc / (k_B T_d) = 0.14 (T_d/10\mbox{ K})^{-1}$ cm is much larger than the typical grain size, we expect $\beta = 2$; detailed grain models show that this expectation holds up to $T_d \approx 150$ K \citep{semenov03a}. \despotic\ leaves both $\sigma_{d,10}$ and $\beta$ as user-settable parameters. A naive expectation is that, at sub-Solar metallicities, $\sigma_{d,10} \propto Z'_d$, where $Z'_d$ is the dust abundance relative to Solar.

The above estimate is valid only as long as the cloud is optically thin to its own cooling radiation, which is true only as long as $\sigma_{d,10} (T_d/10\mbox{ K})^\beta N_{\rm H} \la 1$. Given the small value of $\sigma_{d,10}$ for Milky Way dust, departures from the optically thin regime do not begin until extremely high column densities. However, there are circumstances, for example in the molecular clouds of starburst galaxies, where $T_d$ and $N_{\rm H}$ can be high enough to render the optical depth to cooling radiation large. A truly accurate calculation of the cooling rate in this regime requires a multi-zone numerical treatment with a radiative transfer code such as \texttt{dusty} \citep{ivezic97a} or \texttt{SteinRay} \citep{steinacker03a}, or a sophisticated analytic approximation \citep[e.g.][]{chakrabarti05a}. However, we can obtain a very crude treatment of the optically thick regime by noting that the maximum possible cooling rate for the cloud is simply $\pi R^2 c a T_d^4$, the blackbody rate for a sphere of radius $R = (3/4) N_{\rm H} / n_{\rm H}$ equal to the cloud radius. Rewriting this as a rate per H nucleus, the maximum possible dust cooling rate is
\begin{equation}
\Lambda_{d,\rm thick} = \frac{c a T_d^4}{N_{\rm H}}.
\end{equation}
\despotic\ adopts the approximation
\begin{equation}
\Lambda_d = \min(\Lambda_{d,\rm thin}, \Lambda_{d,\rm thick}).
\end{equation}

\subsubsection{ISRF Heating}

Grains can be heated by absorbing the interstellar radiation field produced by stars. To compute the rate of dust heating from the ISRF, we must perform a calculation similar to that for $\Lambda_d$. In analogy to $\langle Q_\nu(a_g)\rangle_T$, we define $\langle Q_\nu(a_g)\rangle_{\rm ISRF} = \int u_{\nu, \rm ISRF} Q_\nu(a_g) \, d\nu / \int u_{\nu,\rm ISRF}\, d\nu$, where $u_{\nu,\rm ISRF}$ is the energy density of the ISRF at frequency $\nu$, as the ISRF-averaged absorption efficiency. In general $\langle Q_\nu(a_g)\rangle_{\rm ISRF} \gg \langle Q_\nu(a_g)\rangle_T$. Thus, unlike in the case of thermal cooling where optical depth effects are important only in extreme circumstances, attenuation of the ISRF will be important even at modest column densities. As with photoelectric heating, it is clear that there is no single value that describes the rate of dust heating within an optically thick cloud; heating rates will be high at the edge and low at the center. Moreover, unlike in the case of photoelectric heating, the range of photon energies responsible for heating is quite broad, with half the heating coming from photons with wavelengths $>0.31$ $\mu$m even for the unattenuated ISRF (B.~Draine, 2013, priv.~comm.). As a result, the spectrum of the heating field changes as one moves into a cloud and shorter wavelength photons are selectively attenuated. Consequently, in addition to the geometric uncertainty, there is an additional one in the choice of dust cross section to assign. In order to maintain simplicity, \despotic\ does not attempt to treat this problem in detail, but instead uses the same approximation as for photoelectric heating, i.e.\ that the characteristic heating rate is to be computed assuming an attenuation equal to half the mean value for the cloud, using a single grain cross section to compute the attenuation. With this approximation, the heating rate of grains due to the ISRF is
\begin{eqnarray}
\Gamma_{\rm ISRF,thin} & = & \left[\frac{1}{n_{\rm H}} \int \frac{dn}{da_g} \langle Q_{\nu}(a_g)\rangle_{\rm ISRF} \pi a_g^2 \, da_g\right] 
\nonumber \\
& & \qquad {} \cdot c u_{\rm ISRF} e^{-\sigma_{d,\rm ISRF} N_{\rm H}/2} \\
& = & 3.9\times 10^{-24} \chi Z_d' e^{-\sigma_{d,\rm ISRF} N_{\rm H}/2}\mbox{ erg s}^{-1}\mbox{ H}^{-1},
\end{eqnarray}
where $Z'_d$ is the dust abundance relative to the Milky Way value, $u_{\rm ISRF} = \chi u_{\rm MW}$ is the energy density of the ISRF, $u_{\rm MW}$ is the energy density for the Milky Way's ISRF, $\sigma_{\rm ISRF}$ is the cross section we assign for ISRF attenuation, and the numerical coefficient is taken from \citet{goldsmith01a}. The choice of $\sigma_{\rm ISRF}$ is somewhat difficult for the reasons stated above, and if very high accuracy is desired it should be computed on a case-by-case basis. However, a reasonable default for Milky Way dust is $\sigma_{\rm ISRF} = 3\times 10^{-22}$ cm$^2$ H$^{-1}$, which is roughly halfway between the values appropriate for the unextincted ISRF and the value expected for an ISRF extincted by an optical depth of 2 in V band (B.~Draine, 2013, priv.~comm.).

It is worth noting that, because the ISRF is exponentially attenuated by dust, when $\sigma_{d,\rm ISRF} N_{\rm H} \gg 1$ we are likely to find that $\Gamma_{\rm ISRF,thin}$ is negligibly small even when $\chi$ is very large. In this circumstance, the ISRF is so thoroughly attenuated that none of it reaches the cloud interior where we are computing the temperature. However, if this happens, the hot outer parts of the cloud that are directly exposed to the ISRF will heat up and generate a background infrared field within the cloud interior. If the cloud is optically thin to IR cooling radiation the intensity of this field will be low and it can be neglected as a heat source. If the cloud is optically thick to IR, on the other hand, the background IR field will build up, and will heat the cloud interior. \despotic\ provides a mechanism to handle this phenomenon by including an infrared radiation field (see the following section), and in circumstances where ISRF heating is negligible, heating by the infrared radiation field should take its place. As for the case of the cooling rate when the cloud is optically thick to IR, calculating the intensity of the background field in this circumstance requires a more sophisticated model than the one-zone treatment that \despotic\ provides. However, we can solve the limiting case of an extremely optically thick cloud subject to external heating. If such a cloud absorbs all of the background ISRF incident on its surface, the total heating rate is $\pi R^2 c u_{\rm ISRF}$, and the heating rate per H nucleus is
\begin{eqnarray}
\Gamma_{\rm ISRF, thick} & = & \frac{c u_{\rm ISRF}}{N_{\rm H}} 
\nonumber \\
& = & 5.3\times 10^{-25} \frac{\chi}{N_{\rm H,22}}\mbox{ erg s}^{-1}\mbox{ H}^{-1}.
\end{eqnarray}
\despotic\ adopts the approximation
\begin{equation}
\Gamma_{\rm ISRF, thick} = \min(\Gamma_{\rm ISRF,thin}, \Gamma_{\rm ISRF, thick}).
\end{equation}
Equating this with the limiting cooling rate for an extremely opaque cloud, $\Lambda_{d,\rm thick}$, gives an equilibrium temperature for both the dust and the infrared radiation field
\begin{equation}
T_{d,\rm thick} = T_{\rm rad, dust} = \left(\frac{u_{\rm ISRF}}{a}\right)^{1/4} = 2.1\chi^{1/4}\mbox{ K},
\end{equation}
i.e.~the dust and IR radiation field within the cloud reach a temperature such that the radiation energy density within the cloud is equal to the ISRF energy density outside it, as expected for a blackbody.

\subsubsection{Heating by Infrared Radiation and the CMB}

The final source of radiative energy for dust is the background thermal radiation field, and the CMB. Since both of these sources of radiation are thermal, they may be handled using exactly the same mechanics as thermal radiative cooling. The heating rate is therefore
\begin{eqnarray}
\Gamma_{d, \rm IR} & = & \sigma_{d,10} \left(\frac{T_{\rm rad,dust}}{10\mbox{ K}}\right)^\beta c a T_{\rm rad,dust}^4 \\
\Gamma_{d, \rm CMB} & = & \sigma_{d,10} \left(\frac{T_{\rm CMB}}{10\mbox{ K}}\right)^\beta c a T_{\rm CMB}^4.
\end{eqnarray}

\subsubsection{Line Heating}

In addition to emission and absorption of continuum radiation, there are two additional processes that can heat and cool dust grains. The first of these, collisional exchange with the gas, is discussed in \S~\ref{sec:gasheatcool}. The other is absorption of line photons emitted by the gas. If we let
\begin{equation}
\sigma_{d,\nu} = \frac{1}{n_{\rm H}} \int \frac{dn}{da_{\rm g}} \pi a_g^2 Q_\nu \, da_g
\end{equation}
be the population-averaged grain cross section per H nucleus at frequency $\nu$, then the mean optical depth of the cloud to line photons at frequency $\nu$ is $\tau_{d,\nu} = N_{\rm H} \sigma_{d,\nu}$. In principle one could use a detailed grain model to obtain $\sigma_{d,\nu}$ at the frequencies of all the relevant lines. However, this procedure would be cumbersome, and is likely unimportant for most clouds since, not surprisingly, both cooling radiation and observable emission tend to be dominated by lines at frequencies such at clouds are optically thin. Nonetheless, to approximate the effects of clouds becoming optically thick to line radiation, \despotic\ approximates $\sigma_{d,\nu}$ by $\sigma_{d,10} \left(\nu/208\mbox{ GHz}\right)^\beta$, where $\nu$ is the line frequency, and 208 GHz is $(k_B/h)$ multiplied by 10 K. With this approximation, and using the same expression for the line photon escape probability versus optical depth as discussed below in \S~\ref{sec:lines}, we obtain the final heating rate of the dust due to absorption of line photons:
\begin{eqnarray}
\Gamma_{d,\rm line} & = & \sum_{s,ij} (1 - \beta_{d,s,ij}) \Lambda_{s,ij} \\
\beta_{d,s,ij} & = & \frac{1}{1+\frac{3}{8} N_{\rm H} \sigma_{d,10} \left(\nu_{s,ij}/208\mbox{ GHz}\right)^\beta}
\end{eqnarray}
where $\nu_{s,ij}$ is the frequency of the line produced by atoms / molecules of species $s$ transitioning between states $i$ and $j$ (see \S~\ref{sec:lines}), $\beta_{d,s,ij}$ is the escape probability for a photon corresponding to line $ij$ computed using the dust optical depth, and the sum runs over all species $s$ and level pairs $ij$.

\subsection{Level Populations and Line Radiation}
\label{sec:lines}

\subsubsection{Level Populations in Optically Thin Clouds}
\label{sec:levpopthin}

Calculating the line cooling rate requires determining the level populations for all emitting species. Consider an emitting species $s$, and let $E_i$ be the energy of the $i$th quantum state of that species, where the states are numbered by energy so that $E_i < E_{i+1}$ for all states $i$. The degeneracy of state $i$ is $g_i$, and the Einstein coefficient describing the rate of spontaneous radiative transitions from state $i$ to state $j$ is $A_{ij}$, where $A_{ij} = 0$ for $i \leq j$. Finally, let $k_{p, ij}$ be the rate coefficient for collisional transitions from state $i$ to state $j$ induced by collisions with some collision partner $p$; the upward and downward rate coefficients obey the usual relationship $k_{p, ji} = (g_i/g_j) k_{p, ij} \exp(-\Delta E_{ij}/k T_g)$, where $i>j$ and $\Delta E_{ij} = |E_i - E_j|.$ By convention $k_{p,ij} = 0$ for $i=j$.

For our species of interest, we wish to solve for the fraction $f_i$ of atoms / molecules in state $i$, when that species is mixed with a gas of a given bulk composition, number density $n_{\rm H}$, and gas temperature $T_g$, and the cloud is immersed in a sea of cosmic microwave background photons. If the cloud is optically thin to photons at the frequencies of the lines connecting the various states, in statistical equilibrium the various level populations are determined implicitly by the conditions that the rate of transitions into and out of each level balance:
\begin{eqnarray}
\lefteqn{\sum_j f_j \left[q_{ji} + (1 + n_{\gamma,ji}) A_{ji} + \frac{g_i}{g_j} n_{\gamma,ij} A_{ij} \right]}
\nonumber
\\
& = & f_i \sum_k \left[q_{ik} + (1 + n_{\gamma,ik}) A_{ik} + \frac{g_k}{g_i} n_{\gamma,k i} A_{ki}\right],
\label{eq:levpop1}
\end{eqnarray}
where
\begin{eqnarray}
\label{eq:ngamma}
n_{\gamma,ij} & = & \frac{1}{\exp(\Delta E_{ij}/k_B T_{\rm CMB})-1} \\
q_{ij} & = & f_{\rm cl} n_{\rm H} \sum_p x_p k_{p,ij}
\end{eqnarray}
are the photon occupation number at the frequency of the line connecting states $i$ and $j$,\footnote{Naively one would think that, in a cloud that builds up a significant trapped infrared radiation field, then the photon occupation number should also include a contribution from this field, of the same form as equation (\ref{eq:ngamma}) but with $T_{\rm CMB}$ replaced by $T_{\rm rad,dust}$. However, this is often not the case, for the following reason. Even in high column density environments where a significant dust-trapped infrared radiation field builds up, the spectrum of this radiation field is often not Planckian at low frequencies. This is because the dust opacity generally falls as $\nu^2$ at low frequencies, and so even if the dust is opaque to radiation near the peak of the spectral energy distribution, it is usually transparent at low frequencies. This results in a radiation spectrum that is Planckian at higher frequencies but very sub-Planckian at low frequencies, and thus has a much lower photon occupation number that a true blackbody like the CMB. A fully accurate calculation of level populations would account for this effect by solving for the frequency-dependent dust-mediated radiation field and using the appropriate photon occupation number to calculate the level populations. However, as noted above, it is not feasible to determine the dust radiation field accurately in a one-zone model. I therefore choose to optimize the accuracy of \despotic\ for the case of lines at frequencies where the dust is optically thin, since these are, obviously, the lines that are most important for both cooling and observation. This choice dictates that the dust radiation field be ignored when computing the level populations, on the basis that its photon occupation number will be small. However, this choice does limit the accuracy of \despotic\ for lines where infrared pumping is important, as discussed in more detail in \S~\ref{sec:limitations}.} and the rate of collisional transitions between states $i$ and $j$ summed over all collision partners $p$. Here $x_p$ is the abundance of a given collision partner relative to $n_{\rm H}$, and the collision partners considered by \despotic\ are H, He, pH$_2$, oH$_2$, $e$, and H$^+$. As usual, collision rates are multiplied by the clumping factor $f_{\rm cl}$. The left-hand sides of equations (\ref{eq:levpop1}) describe the rate of transitions into state $i$ from all other states $j$, with the first term representing the rate of collisional transitions, the second representing the rate of radiative transitions (including both spontaneous and stimulated emission), and the third term describing the rate of absorptions. The right-hand sides represent the rate of transitions from state $i$ to all other states $k$, with the three terms again representing collisional transitions, spontaneous and stimulated emission, and absorption. These equations are supplemented by the constraint equation
\begin{equation}
\label{eq:fsum}
\sum_i f_i = 1,
\end{equation}
and together equations (\ref{eq:levpop1}) and (\ref{eq:fsum}) constitute a complete system.

For computational purposes it is convenient to rewrite this system as a matrix equation. Consider a species $s$ for which we track $N$ distinct energy levels. With some manipulation, equations (\ref{eq:levpop1}) and (\ref{eq:fsum}) may be rewritten as\footnote{Note that \despotic\ does not use the standard procedure in the stellar atmospheres community of recasting the equations in terms of departure coefficients.This choice is motivated by the fact that, for most of the calculations for which \despotic\ is intended, most of the states of most species will be very far from LTE. This vitiates any advantage to recasting the equations in departure coefficient form.}
\begin{equation}
\label{eq:levpopthin}
\textbfss{M} \bmath{f} = \bmath{b}
\end{equation}
where $\textbfss{M}$ is an $(N+1)\times N$ matrix whose elements are
\begin{eqnarray}
\textbfss{M}_{ij} & = &
- \delta_{ij} + \delta_{i,N+1}
\nonumber\\
& & 
{}+\frac{q_{ji} + (1 + n_{\gamma,ji}) A_{ji} + \frac{g_i}{g_j} n_{\gamma,ij} A_{ij}}
{\sum_k \left[q_{ik} + (1 + n_{\gamma,ik})A_{ik} + \frac{g_k}{g_i} n_{\gamma,ki} A_{ki}\right]}
,
\end{eqnarray}
$\bmath{b}$ is a vector of length $N+1$ whose elements are
\begin{equation}
\bmath{b}_i = \delta_{i,N+1}
\end{equation}
and $\bmath{f}$ is a vector of length $N$ whose elements are the fractional level populations $f_i$. By convention $q_{ij} = A_{ij} = n_{\gamma,ij} = 0$ for $i = N+1$ or $j=N+1$. The off-diagonal elements of matrix $\textbfss{M}$ in rows $i\leq N$ have a simple physical meaning: element $ij$ is the rate coefficient for transitions (adding both radiative and collisional processes) into state $i$ from state $j$, normalized by the sum of the rate coefficients for all transitions out of state $i$ to any other state. The final row of $\textbfss{M}$, $i=N+1$, implements the constraint equation that the sum of all fractional level populations is unity.

\begin{figure}
\includegraphics[width=84mm]{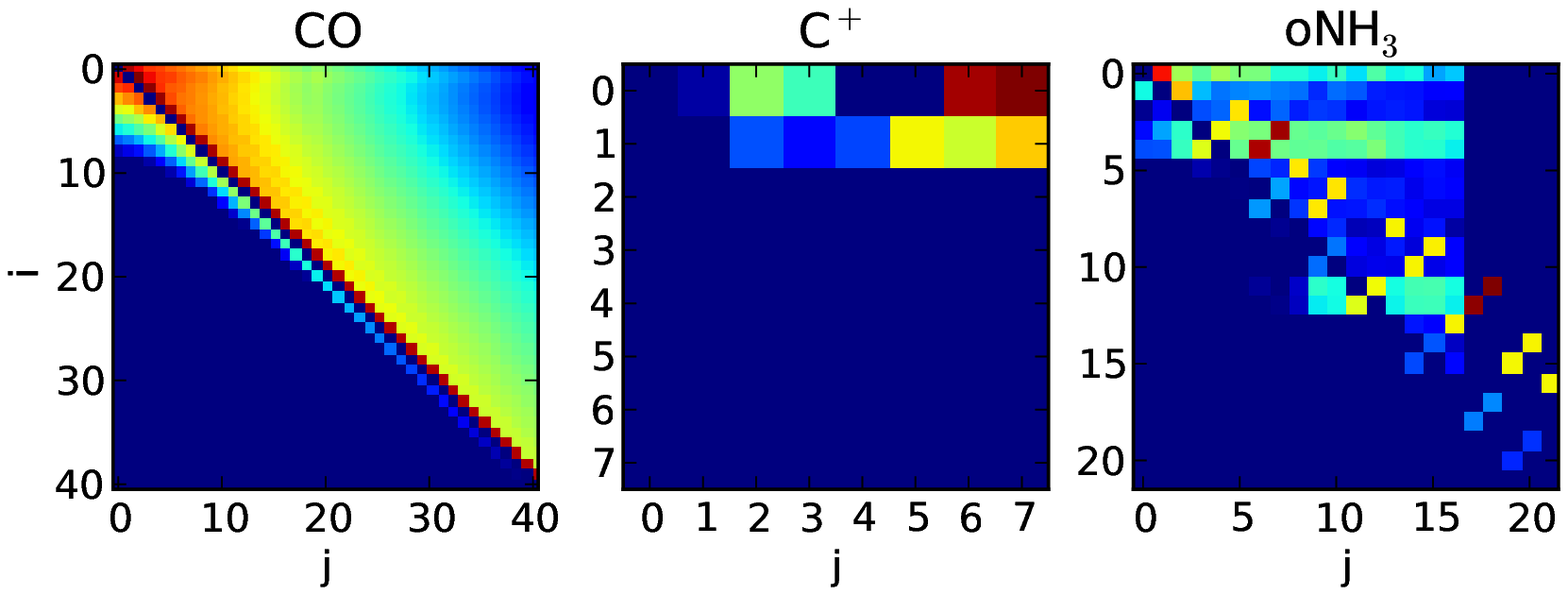}
\caption{
\label{fig:matrix1}
A graphical representation of the matrices $\textbfss{M}$ calculated for CO, C$^+$, and oNH$_3$ using the conditions described in Appendix \ref{sec:levpopthin}. The color of each block represents the value of the corresponding element $ij$ of $\textbfss{M}$, excluding the elements with $i=N+1$, which are all unity. The color scale is normalized and logarithmic, so that the largest element of $\textbfss{M}$ is shown in red, while dark blue corresponds to a value of $10^{-15}$ times the value of the largest element. Values on the diagonal are masked, since they are negative. Recall that the value of an element $M_{ij}$ is the sum of the rate coefficients describing transitions into state $i$ from state $j$ (including both collisional and radiative transitions), normalized by the sum of the rate coefficients out of state $i$ into any other state. Thus elements above the diagonal represent downward transitions, while those below the diagonal are upward transitions. 
}
\end{figure}

Robust numerical solution of equation (\ref{eq:levpopthin}) requires considerable care, because when the transition probabilities into certain states are very low, the matrix $\textbfss{M}$ can by extremely ill-conditioned, making accurate numerical solution impossible. Figure \ref{fig:matrix1} illustrates the nature of the problem by graphically displaying $\textbfss{M}$ for the species CO, C$^+$,\footnote{LAMDA offers two data tables for C$^+$, one including only the low-lying fine-structure levels, and one also including the higher energy levels connected to them by UV lines. For the purposes of this example, I use the data file including the UV levels.} and oNH$_3$, all computed for a cloud with $n_{\rm H} = 10^3$ cm$^{-3}$ composed of pure pH$_2$ at a temperature of 10 K, embedded in the cosmic microwave background at temperature $T_{\rm CMB} = 2.73$ K. For simplicity the cloud is assumed to be optically thin and to have a clumping factor $f_c = 1$. The matrix describing CO has a wide range of transition rates, but every row and column contains at least one transition rate coefficient whose magnitude is comparable to that of the largest elements of the matrix. As a result, the CO matrix is well-conditioned: for the example shown in Figure \ref{fig:matrix1}, the condition number is 135. This presents no challenges for numerical solution. On the other hand, the matrices for C$^+$ and oNH$_3$ both have the property that a few elements are much larger than most of the rest of the matrix. As a result they have condition numbers of $2.7\times 10^{36}$ and $1.2\times 10^{13}$, respectively. These condition numbers imply that a numerical solution to equation (\ref{eq:levpopthin}) for C$^+$ and oNH$_3$ would have $\sim 36$ and $\sim 13$ fewer digits of accuracy than machine accuracy, rendering numerical solutions obtained for these matrices meaningless.

The high condition numbers of the matrices are a direct result of the physical processes they describe, and the divergence in timescales between transitions between different states. In the matrix for C$^+$, for example, the largest elements correspond to transitions between the $^2$S$_{1/2}$ and $^2$D$_{3/2}$ states (states 7 and 6, respectively, in Figure \ref{fig:matrix1}) and the ground, $^2$P$^o_{1/2}$ state (state 0 in Figure \ref{fig:matrix1}). These have Einstein coefficients $A\sim 10^9$ s$^{-1}$, compared with the fine-structure transition between the first two states, which has $A = 2.3\times 10^{-6}$ s$^{-1}$. Similarly, for oNH$_3$, the largest elements of $\textbfss{M}$ describe transitions such as $(J,K)_v = (7,6)_1 \rightarrow (6,6)_0$ and $(7,6)_0\rightarrow(6,6)_1$ (elements $ij=11,18$ and $12,17$, respectively, in Figure \ref{fig:matrix1}), with $A \sim 0.1$ s$^{-1}$, while the inversion transitions that are most commonly observed (e.g.~$(6,6)_1\rightarrow (6,6)_0$, element 11,12), have $A\sim 10^{-7}$ s$^{-1}$.

\begin{figure}
\includegraphics[width=84mm]{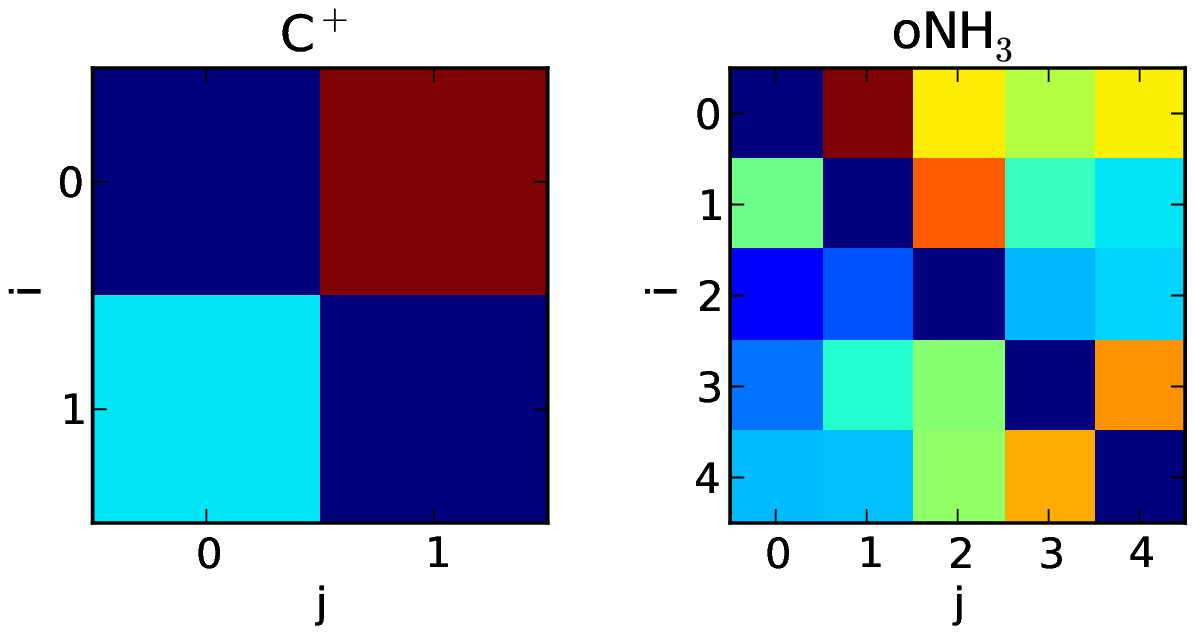}
\caption{
\label{fig:matrix2}
Same as Figure \ref{fig:matrix1} for C$^+$ and oNH$_3$, after level reduction as described in the text.
}
\end{figure}

\despotic\ handles the task of solving equation (\ref{eq:levpopthin}) as follows. First it constructs the matrix $\textbfss{M}$ from the specified cloud properties. It then checks the condition number of $\textbfss{M}$. If it is acceptably small, \despotic\ then solves the equation using the \texttt{LAPACK} routine \texttt{lstsq}. If the condition number is excessively large, \despotic\ employs two strategies to reduce it before calling \texttt{lstsq}. First, in many cases high condition numbers are associated with large rates for downward transitions from high-energy levels. In the simple one-zone statistical equilibrium model used by \despotic, the population of any level will be bounded between the values expected when the atom is in LTE at $T_g$ and when it is in LTE at $T_{\rm CMB}$. \despotic\ calculates these two limiting values, and if it finds that they are below a numerical floor\footnote{\despotic\ sets this floor equal to the machine epsilon value for the platform on which it is operating, which is usually $\sim 10^{-15}$.}, it simply sets the populations of those levels to the floor, and removes the associated rows and columns from matrix $\textbfss{M}$. If these rows and columns contain large elements, the condition number of the matrix is likely to be reduced. For the examples shown in Figure \ref{fig:matrix1}, applying this procedure eliminates the 6 highest energy levels for C$^+$ (and thus the six bottom- and right-most rows and columns in $\textbfss{M}$) and the 11 highest energy levels for oNH$_3$. In turn, this reduces the condition number for the C$^+$ matrix to $7.5\times 10^4$, low enough to allow numerical solution with tolerable accuracy. Figure \ref{fig:matrix2} shows the same graphical representation of the matrix for C$^+$ as in Figure \ref{fig:matrix1} after this level reduction procedure.

Unfortunately this procedure alone still leaves the matrix for oNH$_3$ with an unacceptably-high condition number of $7.6\times 10^{11}$. This is because not all of the large matrix elements for this case apply only to the high-energy levels. The levels $(3,3)_0$, $(3,3)_1$, $(4,3)_0$, and $(4,4)_0$ are close enough to the ground state in energy for their populations not to be entirely negligible, but they still have Einstein $A$ values far larger than those associated with the inversion transitions. 

The second strategy \despotic\ employs is to eliminate levels whose populations will be small because the rate coefficients for transitions out of them greatly exceeds the rate coefficients for transitions into them. Specifically, consulting equation (\ref{eq:levpop2}), the total rate at which a particle in state $i$ will transition to another state is
\begin{equation}
\Gamma_{i,\rm out} = \sum_{k} \left[q_{ik} + \beta_{ik} \left(1 + n_{\gamma,ik}\right)A_{ik} + \beta_{ki} \frac{g_k}{g_i} n_{\gamma,ki} A_{ki}\right].
\end{equation}
The total rate of transitions into state $i$ is given by the left-hand side of equation (\ref{eq:levpop2}), and since each $f_j$ is strictly less than unity, the rate of transitions into state $i$ is bounded above by
\begin{equation}
\Gamma_{i,\rm in} < \sum_j  \left[q_{ji} + \beta_{ji} \left(1 + n_{\gamma,ji}\right)A_{ji} + \beta_{ij} \frac{g_i}{g_j} n_{\gamma,ij} A_{ij}\right].
\end{equation}
Thus the population of state $i$ is strictly bounded above by
\begin{equation}
f_i < f_{i,\rm lim} \equiv \frac{\Gamma_{i,\rm in}}{\Gamma_{i,\rm out}}.
\end{equation}
For ill-conditioned matrices, this ratio is very small for some levels and very large for others. In the example of oNH$_3$, once the high-temperature levels have been eliminated, the value of $f_{i,\rm lim}$ runs from a minimum of $1.3\times 10^{-7}$ to a maximum of $1.2\times 10^5$. Thus if $\textbfss{M}$ remains ill-conditioned after the high-temperature levels have been eliminated, \despotic\ finds the level with the smallest value of $f_{i,\rm lim}$ and eliminates it in exactly the same manner as the high temperature levels. If necessary it repeats this procedure with the next smallest value of $f_{i,\rm lim}$ and  so forth, until the condition number of the matrix is acceptably small. Figure \ref{fig:matrix2} shows the matrix for oNH$_3$ after this procedure is complete, leading to the elimination of all levels by the five lowest energy ones. The condition number of the resulting matrix is $3.5\times 10^5$, again allowing accurate numerical solution.

\subsubsection{Level Populations in Optically Thick Clouds}

If the cloud is optically thick, these equations must be modified to account for the fact the effects of the trapped radiation field that builds up inside the cloud. To handle this case, \despotic\ uses the standard escape probability approximation, in which the level populations are assumed to be uniform, and every transition $ij$ is assigned an escape probability $\beta_{ij}$, which gives the volume-averaged probability that, when an atom or molecule radiatively decays from state $i$ to state $j$, the associated photon will escape from the cloud rather than being resonantly absorbed within it. With this approximation, the modified equations simply become  \citep[e.g.][]{draine11a}
\begin{eqnarray}
\lefteqn{\sum_j f_j \left[q_{ji} + \beta_{ji} (1 + n_{\gamma,ji}) A_{ji} + \beta_{ij} \frac{g_i}{g_j} n_{\gamma,ij} A_{ij} \right]}
\nonumber
\\
& = & f_i \sum_k \left[q_{ik} + \beta_{ik} (1 + n_{\gamma,ik}) A_{ik} + \beta_{ki} \frac{g_k}{g_i} n_{\gamma,k i} A_{ki}\right].
\label{eq:levpop2}
\end{eqnarray}

The escape probability may be computed using several possible approximations, which are appropriate for different cloud geometries. By default, \despotic\ uses the approximate result from \citet{draine11a} for uniform spherical clouds,
\begin{eqnarray}
\label{eq:beta}
\beta_{ij} & = & \frac{1}{1 + \frac{3}{8} \tau_{ij}} \\
\label{eq:tauij}
\tau_{ij} & = & \frac{g_i}{g_j} \frac{A_{ij} \lambda_{ij}^3}{4 (2\pi)^{3/2} \sigma_{\rm tot}} x_s N_{\rm H} f_j \left(1 - \frac{f_i g_j}{f_j g_i}\right),
\end{eqnarray}
where $\tau_{ij}$ is the optical depth corresponding to a column $N_{\rm H}$, $\lambda_{ij} = hc/\Delta_{ij}$ is the wavelength of transition $ij$, $\sigma_{\rm tot} = \sqrt{\sigma_{NT}^2 + c_s^2/\mu_s}$, $\mu_s$ is the molecular weight of the emitting species in units of $m_{\rm H}$, and $x_s$ is the abundance of the emitting species per H nucleus. Note that, in the expression for $\beta_{ij}$, the coefficient on $\tau_{ij}$ differs by a factor of $(3/4)$ from that given in \citet{draine11a} because \citeauthor{draine11a} defines $\tau_{\rm ij}$ using the center-to-edge rather than the mean column density. 

The code can also use one of two other approximations. For a slab geometry, the escape probability is \citep{de-jong75a}
\begin{equation}
\label{eq:betaslab}
\beta_{ij, \rm slab} = \frac{1-e^{-3\tau_{ij}}}{3\tau_{ij}},
\end{equation}
where $\tau_{ij}$ is again given by equation \ref{eq:tauij}, but now $N_{\rm H}$ is interpreted as the column density of the slab rather than the mean column density of a sphere. Finally, \despotic\ can use the large velocity gradient (LVG) approximation, in which the escape probability is computed from a Sobolev approximation and the geometry is therefore irrelevant \citep{de-jong80a}. In this case
\begin{eqnarray}
\label{eq:betaLVG}
\beta_{ij, \rm LVG} & = & \frac{1 - e^{-\tau_{ij,\rm LVG}}}{\tau_{ij,\rm LVG}} \\
\tau_{ij,\rm LVG} & = & \frac{g_i}{g_j} \frac{A_{ij} \lambda_{ij}^3}{8 \pi |dv_r/dr|} x_s N_{\rm H} f_j \left(1 - \frac{f_i g_j}{f_j g_i}\right).
\end{eqnarray}

For whichever choice of geometry, the above equations determine $\beta_{ij}$ in terms of $f_i$ and other known quantities, and together with equations (\ref{eq:fsum}) and (\ref{eq:levpop2}) they again form a complete system that may be solved for $f_i$. In the optically thin limit, $\beta_{ij} \rightarrow 1$ for all $ij$, and equations (\ref{eq:levpop2}) reduce to equations (\ref{eq:levpop1}).

\despotic\ solves this system numerically via the following procedure. For specified escape probabilities $\beta_{ij}$, the procedure for calculating the level populations is identical to that given in Appendix \ref{sec:levpopthin} for optically thin clouds, except that matrix $\textbfss{M}$ becomes
\begin{eqnarray}
\lefteqn{
\textbfss{M}_{ij} =
- \delta_{ij} + \delta_{i,N+1}
}
\nonumber \\
& & 
{}+\frac{q_{ji} + \beta_{ji} (1 + n_{\gamma,ji}) A_{ji} + \beta_{ij} \frac{g_i}{g_j} n_{\gamma,ij} A_{ij}}
{\sum_k \left[q_{ik} + \beta_{ik} (1 + n_{\gamma,ik})A_{ik} + \beta_{ki} \frac{g_k}{g_i} n_{\gamma,ki} A_{ki}\right]}.
\label{eq:matrixthick}
\end{eqnarray}
However, the escape probabilities $\beta_{ij}$ are not known in advance, and they and the level populations must instead be computed iteratively. Again, by convention, $\beta_{ij} = 0$ for $i=N+1$.

Let $f_i^{(n)}$ be the current best guess for the level populations after $n$ iterations. At every step of the iteration, \despotic\ uses $f_i^{(n)}$ to compute a new estimate for the escape probabilities $\beta_{ij}^{(n)}$ of all lines from equation (\ref{eq:beta}), (\ref{eq:betaslab}), or (\ref{eq:betaLVG}), constructs the matrix $\textbfss{M}$ following equation (\ref{eq:matrixthick}) using $\beta_{ij}^{(n)}$, and then solves equation (\ref{eq:levpopthin}) to obtain a new set of level population estimate $f_i^{(*)}$. \despotic\ then checks if the level populations have converged by computing the absolute and relative residuals
\begin{eqnarray}
\mbox{abs.~resid.} & = &\max_i |f^{(n)}_{i} - f^{(*)}_{i}| \\
\mbox{rel.~resid.} & = & \max_i \frac{|f^{(n)}_{i} - f^{(*)}_{i}|}{ \max(f_i^{(n)}, f_i^{(*)})}
\end{eqnarray}
and comparing them to specified tolerances. If the residuals exceed the specified tolerances, \despotic\ generates a new set of level populations
\begin{equation}
f_i^{(n+1)} = D f_i^{(*)} + (1 - D) f_i^{(n)},
\end{equation}
where the damping factor $D$ is in the range $(0,1]$. Larger values of $D$ represent more aggressive attempts to converge to the solution rapidly, at the cost of a higher risk of non-convergence. \despotic\ chooses a default $D=0.5$, but this value can be altered by the user, and in calculations where level populations are likely to be computed repeatedly (for example when computing thermal equilibria), \despotic\ catches non-convergences automatically and attempts to recompute using a smaller value of $D$, thereby preventing the entire computation from being derailed.

To start the process, \despotic\ initializes by setting $f_i^{(0)}$ to either the currently-stored level populations for a given cloud or, if none are available, their LTE values for the gas temperature. Initializing to the currently-stored level populations ensures that, when level populations must be computed repeatedly under physical conditions that very only slightly, as in many of the examples given in \S~\ref{sec:examples}, the initial guess will be close to the correct level populations, and convergence will be rapid.

Once a converged solution for the level populations is found, \despotic\ calculates the line luminosities using equation (\ref{eq:linelum}), and the integrated intensity and brightness temperature emerging from the cloud via equations (\ref{eq:intintensity}) and (\ref{eq:TBint}). This is identical to the optically thin case, except that the escape probabilities $\beta_{ij}$ may not be unity.

\subsubsection{Line Cooling Rates and Intensities}

Given a set of level populations determined by solving the equations given in the previous section, the cooling rate of the cloud due to emission by line $ij$ of species $s$ is given by
\begin{equation}
\label{eq:linelum}
\Lambda_{s,ij} = \beta_{ij} \left[(1 + n_{\gamma,ij}) f_i - \frac{g_i}{g_j} n_{\gamma,ij} f_j\right] A_{ij} \Delta E_{ij} x_s.
\end{equation}
Note that this is the net cooling rate, in that the first term in brackets represents the rate of spontaneous plus stimulated emission per emitting atom / molecule, while the second term is the rate of absorption of background photons. Thus $\Lambda_{s,ij}$ is the rate of energy loss via line emission minus the rate of energy gain from absorption of the background radiation field. If $T_g < T_{\rm CMB}$, then $\Lambda_{s,ij}$ will be negative, indicating a net gain in energy. The total cooling rate $\Lambda_s$ for species $s$ is simply the sum over all level pairs,
\begin{equation}
\Lambda_s = \sum_{ij} \Lambda_{s,ij}.
\end{equation}

The emergent frequency-integrated intensity $I_{s,ij}$ and velocity-integrated brightness temperature $T_{B,s,ij}$ for each line are related to the cooling rate via
\begin{eqnarray}
\label{eq:intintensity}
I_{s,ij} & = & \frac{\beta_{d,s,ij}}{4\pi} \Lambda_{s,ij} N_{\rm H} \\
\label{eq:TBint}
T_{B,s,ij} & = & \lambda_{s,\ij} \frac{h/k_B}{\ln \left[1 + 2 h \nu_{s,ij}^3 /c^2 I_{s,ij}\right]}.
\end{eqnarray}
Note the factor $\beta_{d,s,ij}$ in equation (\ref{eq:intintensity}), which accounts for absorption of line radiation by dust internal to the cloud. This effect need not be included when calculating the level populations, under the assumption that any line photon absorbed by dust will not be re-emitted in resonance with the line and thus cannot cause an absorption elsewhere in the cloud. This assumption is well-justified for the infrared and radio lines for which \despotic\ is specialized, since absorption opacities exceed scattering opacities at these frequencies by many orders of magnitude. However, dust absorption must still be included when calculating the observable intensity emerging from the cloud, since photons absorbed by dust will ultimately be emitted as thermal continuum rather than lines.

\section{Calculation of Line Shapes}
\label{app:lineshape}

Consider a line of sight passing through a spherical cloud at an offset distance $d$ from the cloud center (see Figure \ref{fig:cloudfig}), and let $n_s$, $T$, $v$, and $\sigma_{\rm NT}$ be the number density of the emitting species $s$, gas temperature, the radial velocity, and the non-thermal velocity dispersion. Each of these in general can be functions of $r$. Now consider a spectral line of this species connecting an upper state $u$ to a lower state $\ell$. Under the assumption of LTE, number densities of species $s$ in the upper and lower states are
\begin{equation}
n_{\ell} = \frac{g_\ell e^{-E_\ell/k_B T}}{Z_s(T)} n_s, \qquad
n_u = \frac{g_u e^{-E_u/k_B T}}{Z_s(T)} n_s,
\end{equation}
where $g_i$ is the degeneracy of state $i$, $E_i$ is the energy of the state, and $Z_s(T)$ is the partition function for species $s$ at temperature $T$.

The equation of radiative transfer along the chosen line of sight reads
\begin{equation}
\frac{dI_\nu}{ds} = j_\nu - \kappa_\nu I_\nu,
\end{equation}
where $s$ is the position along the line of sight, defined such that $s=0$ is the cloud midplane, and integration through the cloud proceeds from $s=-\sqrt{R^2-d^2}$ to $+\sqrt{R^2-d^2}$ (Figure \ref{fig:cloudfig}). The emission and absorption coefficients are given by
\begin{eqnarray}
\kappa_\nu & = & \frac{g_u}{g_\ell} n_\ell \frac{\lambda^2}{8\pi} A_{u\ell} \phi_\nu \\
j_\nu & = & \kappa_\nu B_\nu(T)
\end{eqnarray}
where $\lambda = hc/\Delta E$ is the wavelength of the transition, $\Delta E = E_u-E_\ell$ is the energy difference between the levels, $A_{u\ell}$ is the Einstein coefficient for the transition, and $\phi_\nu$ is the line profile. This is given by
\begin{equation}
\phi_\nu = \frac{1}{\sqrt{2\pi \sigma_\nu^2}} \exp\left[-\frac{(\nu-\nu_0)^2}{2\sigma_\nu^2}\right],
\end{equation}
where $\sigma_\nu$ and $\nu_0$ are the dispersion in frequency and the frequency of line center, given by
\begin{eqnarray}
\sigma_\nu & = & \frac{1}{\lambda} \sqrt{\sigma_{\rm NT}^2 + \frac{k_B T}{\mu_s m_{\rm H}}} \\
\nu_0 & = & \frac{\Delta E}{h} \left(1 - \frac{v}{c} \sin\frac{s}{\sqrt{s^2+d^2}}\right),
\end{eqnarray}
where $\mu_s$ is the mass of a particle of species $s$, measured in H masses. The transfer equation may be non-dimensionalized via the change of variables. We let $x=s/R$ be the dimensionless position, $f=\nu/(\Delta E/h)$ be the dimensionless frequency, $\mathcal{I}_f = I_\nu/(A_{u\ell} n_s(R) h R)$ be the dimensionless intensity, and we normalize all the position-dependent quantities to their values at the cloud edge: $n_s' = n_s/n_s(R)$, $t = T/T(R)$, $\psi = \sigma_{\rm NT}/\sigma_{\rm NT}(R)$, and $u=v/v(R)$. With these definitions, after some manipulation the transfer equation becomes
\begin{equation}
\label{eq:transfer}
\frac{d\mathcal{I}_f}{dx} = n'_s\frac{g_u e^{-\Theta_\ell/t}}{4\pi Z_s(T)} \left[e^{-\Theta/t} - \tau_0 \left(1-e^{-\Theta/t}\right) \mathcal{I}_f \right] \phi_f,
\end{equation}
where
\begin{eqnarray}
\phi_f & = & \frac{1}{\sqrt{2\pi \sigma_f^2}} \exp\left[-\frac{(f-f_0)^2}{2\sigma_f^2}\right] \\
f_0 & = & 1 - \beta u \sin\frac{x}{\sqrt{x^2+(d/R)^2}} \\
\sigma_f & = & \sqrt{\beta_s^2 t + \beta_\sigma^2 \psi^2}
\end{eqnarray}
and we have defined the dimensionless ratios
\begin{eqnarray}
\label{eq:lineprofstart}
\Theta_\ell = \frac{E_\ell}{k_B T(R)} & \quad &
\Theta = \frac{E_u-E_l}{k_B T(R)} \\
\beta = \frac{v(R)}{c} & &
\beta_s = \frac{k_B T(R)}{\mu_s m_{\rm H} c} \\
\beta_\sigma = \frac{\sigma_{\rm NT}(R)}{c} & &
\tau_0 = \frac{A_{u\ell} \lambda^3 n_s(R) R}{2 c}.
\label{eq:lineprofend}
\end{eqnarray}
The dimensionless intensity $\mathcal{I}_f$ at any dimensionless frequency $f$ may be obtained by integrating equation (\ref{eq:transfer}) from $x=-\sqrt{1-(d/R)^2}$ to $x = +\sqrt{1-(d/R)^2}$ subject to the boundary condition $\mathcal{I}_f = B_\nu(T_{\rm CMB})/(A_{u\ell} n_s(R) h R)$ at the lower limit of integration.

\section{Comparison Between \despotic\ and \texttt{RADEX}}
\label{sec:radex}

As a check on \despotic\ and to illustrate its strengths and weaknesses, in this Appendix I provide a detailed comparison between \despotic\ and \texttt{RADEX} \citep{van-der-tak07a}. \texttt{RADEX} does not include \despotic's capabilities for computing heating and cooling rates, thermal equilibria, time-dependent thermal evolution, or line shapes, so this test is limited to the capabilities that the two codes have in common: computing level populations and emergent line intensities from a cloud of specified physical properties.

For the purposes of this test, I compute the CO spectral line energy distribution for a cloud with temperature of $T_g = 10$ K, a full-width-at-half-maximum velocity spread of 2.0 km s$^{-1}$, and a CO abundance $x_{\rm CO} = 10^{-4}$ over a grid of volume densities from $n_{\rm H} = 10^2 - 10^8$ cm$^{-3}$ and column densities $N_{\rm H} = 10^{14} - 10^{24}$ cm$^{-2}$, in steps of 0.2 dex in both dimensions. The grid is chosen to cover a wide range of conditions, from optically thin to optically thick, and from thermalized (for the first for levels) to highly sub-thermal. For both codes the background radiation field is set to the CMB value of 2.73 K, and I use slab geometry for the escape probability calculation, since \texttt{RADEX} and \despotic\ use the same approximate expression for the escape probability in that case. I perform the \texttt{RADEX} computation using a slightly modified version of the \texttt{radex\_grid.py} wrapper that is distributed as part of the \texttt{RADEX} package.

To ensure that the computations are identical, for the \despotic\ calculation I set the non-thermal velocity dispersion to a temperature-dependent value $\sigma_{\rm NT}$ in \despotic\ to $\sigma_{\rm NT} = [\mbox{FWHM}^2/8\ln 2 + c_s^2/\mu_{\rm CO} m_{\rm H}]^{1/2}$, where $\mu_{\rm CO} = 28$ is the molecular weight of CO. This guarantees that the velocity dispersions are the same in the two calculations. Similarly, I disable clumping and I set all dust opacities to 0 in \despotic, since \texttt{RADEX} includes neither clumping nor dust absorption. Finally, I set the abundances of ortho- and para-H$_2$ in \despotic\ to
\begin{eqnarray}
x_{\rm oH_2} & = & \frac{9 e^{-2\theta_{\rm rot}/T_g}}{1 + 9 e^{-2\theta_{\rm rot}/T_g}} \\
x_{\rm pH_2} & = & \frac{1}{1 + 9 e^{-2\theta_{\rm rot}/T_g}},
\end{eqnarray}
consistent with \texttt{RADEX}'s hardwired assumption that the ratio of ortho- to para-H$_2$ is given by the thermal ratio of the populations of the H$_2$ $J=1$ to $J=0$ states.

Comparison of the results indicates that the level populations and line optical depths returned by the two codes are identical to the level of precision with which \texttt{RADEX} writes output. The line fluxes returned by the codes, interestingly enough, are not identical, and this is due to a minor lack of self-consistency in the escape probability approximation itself. To compute the frequency-integrated line flux, \despotic\ first computes the total rate of energy emission per H nucleus from equation (\ref{eq:linelum}), and then computes the integrated intensity and brightness temperature from equations (\ref{eq:intintensity}) and (\ref{eq:TBint}). In contrast, \texttt{RADEX} computes the output intensity using the transfer equation for a uniform medium. It uses the level populations to compute an excitation temperature $T_{{\rm ex},ij}$ between every pair of levels $i$, $j$, computes the optical depth at line center $\tau_{ij}$ from the level populations (equation 21 of \citealt{van-der-tak07a}), and then computes the emergent integrated intensity as
\begin{equation}
\int \left[B_\nu(T_{{\rm ex},ij}) \left(1-e^{-\tau_{ij}}\right) + e^{-\tau_{ij}} B_\nu(T_{\rm CMB})\right] \phi_\nu \, d\nu,
\end{equation}
where $\phi_\nu$ is the line shape function, which is taken to be a Gaussian whose dispersion is determined by the input FWHM. To obtain the cooling rate per H nucleus, this quantity is simply divided by the total column density. In the limit of high optical depth, and neglecting the contribution of the background radiation field (which is indeed negligible in the example given), with some algebra one may show that \texttt{RADEX}'s expression reduces to
\begin{equation}
\label{eq:coolrateradex}
\Lambda_{s,ij} = \frac{1}{\tau_{ij}} A_{ij} \Delta E_{ij} f_i,
\end{equation}
while \despotic's expression (equation \ref{eq:linelum}) reduces to
\begin{equation}
\Lambda_{s,ij} = \beta_{ij} A_{ij} \Delta E_{ij} f_i.
\end{equation}
In the optically thin limit, the factors of $1/\tau_{ij}$ and $\beta_{ij}$ are omitted, rendering the expressions identical. As expected, in the optically thin limit the two codes produce results that are identical to the precision with which \texttt{RADEX} writes output. For optically thick lines, on the other hand, the two expressions above are identical only if $\beta_{ij} \rightarrow 1/\tau_{ij}$ as $\tau_{ij}\rightarrow \infty$. This is the case for the LVG approximation, and for the expression \texttt{RADEX} uses for spherical geometry\footnote{In spherical geometry one must be careful to correct for the fact that \texttt{RADEX} defines the optical depth appearing in equation (\ref{eq:coolrateradex}) as measured along a cloud diameter, which is larger than the projection-averaged optical depth used in \despotic\ by a factor of $3/2$.}, but it is not true for slab geometry or for the approximation that \despotic\ uses in spherical geometry.

This disagreement arises from a fundamental limitation of the escape probability approximation. In this approximation, one assumes that there is a uniform escape probability that characterizes the entire cloud, and that the level populations within the cloud are also uniform, but these two assumptions are not fully consistent. \despotic's calculation of the line luminosity follows from the former assumption, while \texttt{RADEX}'s follows from the latter. However, the former assumption is preferable for the types of problems that \despotic\ is intended to solve, because it enforces strong consistency between the rate of photon emission and escape from the cloud, and the rate of rate of energy loss via line cooling.

Finally, I note that, in timing tests, \texttt{RADEX} performs this calculation a factor of $\sim 5$ faster than \despotic. This difference is not surprising, given that \texttt{RADEX} is a single-purpose tool written in Fortran, compiled with heavy optimization, and where many decisions are made at compile time (e.g.~the geometry used to compute escape probabilities), while \despotic\ is a much more general-purpose and interactive tool written in a non-optimized language, and with a large number of options that are specified at run-time. In neither case is the computational cost prohibitive, however. On the workstation where I performed the tests, the full grid of 1581 models required roughly 35 seconds to evaluate for \texttt{RADEX}, and a bit under 3 minutes for \despotic.

\end{document}